# Phase field simulations of thermal annealing for all-small molecule organic solar cells

Yasin Ameslon[1], Olivier J. J. Ronsin[1], Christina Harreiß[2], Johannes Will[2], Stefanie Rechberger[2], Mingjian Wu[2], Erdmann Spiecker[2] and Jens Harting[1,3]

[1]Helmholtz Institute Erlangen-Nürnberg for Renewable Energy, Forschungszentrum Jülich, Fürther Strasse 248, 90429 Nürnberg, Germany

[2]Institute of Micro- and Nanostructure Research (IMN) & Center for Nanoanalysis and Electron Microscopy (CENEM), Interdisciplinary Center for Nanostructured Films (IZNF), Department of Materials Science and Engineering, Friedrich-Alexander-Universität Erlangen-Nürnberg, Cauerstrasse 3, 91058 Erlangen, Germany

[3]Department of Chemical and Biological Engineering and Department of Physics, Friedrich-Alexander-Universität Erlangen-Nürnberg, Cauerstrasse 1, 91058 Erlangen, Germany

# Abstract

Interest in organic solar cells (OSCs) is constantly rising in the field of photovoltaic devices. The device performance relies on the bulk heterojunction (BHJ) nanomorphology, which develops during the drying process and additional post-treatment. This work investigates the effect of thermal annealing (TA) on the all-small molecule DRCN5T: $PC_{71}BM$ blend with phase field simulations. The objective is to determine the physical phenomena driving the evolution of the BHJ morphology for a better understanding of the post-treatment/morphology relationship. Phase-field simulation results are used to investigate the impact on the final BHJ morphology of the DRCN5T crystallization-related mechanisms, including nucleation, growth, crystal stability, impingement, grain coarsening, and Ostwald ripening, of the amorphous-amorphous phase separation (AAPS), and of diffusion limitations. The comparison of simulation results with experimental data shows that the morphological evolution of the BHJ under TA is dominated by dissolution of the smallest, unstable DRCN5T crystals and anisotropic growth of the largest crystals.

## Keywords

# Table of Contents

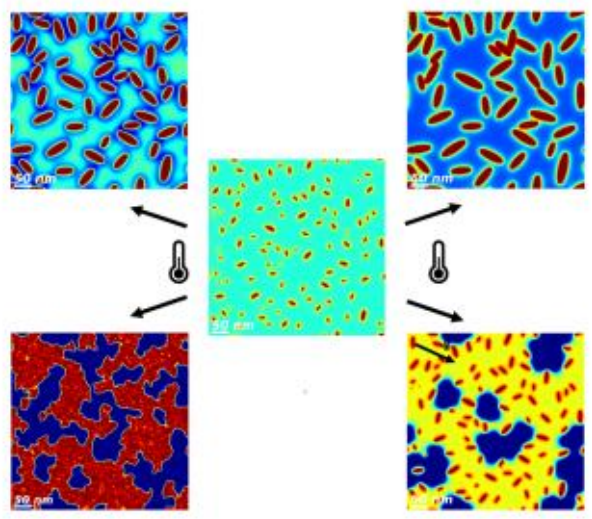

*Figure 1: Simulated DRCN5T:$PC_{71}BM$ morphologies under various thermal annealing conditions. The centered image corresponds to the as-cast simulated film, the other images correspond to the morphology after thermal annealing under diffusion limited (top left corner) and normal crystal growth (top right corner) regimes, with nucleation (bottom left corner) and with amorphous-amorphous phase separation (bottom right corner).*

# 1. Introduction

Organic solar cells (OSC) have the potential to become the cheapest electricity source, even cheaper than silicon based solar cells. This potential comes from the ability to coat large areas quickly and at very low costs using printing methods.[1,2] In addition, they represent an excellent option for fabricating flexible, light-weight, and semi-transparent solar cells, which opens the way to applications for building integrated photovoltaics, agrivoltaics or portable electronics.[2–4] In order to ensure charge carrier separation, OSC absorber layers are composed of two materials, one electron donor and one electron acceptor, which might be polymers or small molecules. Recently, the development of non-fullerene acceptor small molecules, particularly Y6 and its derivatives, led to OSCs reaching 19 % power conversion efficiency (PCE) with polymer donors.[5] On the other hand, OSC with a small-molecule donor, the so-called all-small molecule (ASM) solar cells have reached recently PCE around 18%.[6] As compared to polymer OSC, ASM OSC still suffer from slightly lower PCE, shorter lifetime and scalability problems due to the need for posttreatment methods such as thermal annealing (TA) or solvent vapor annealing (SVA) to obtain high PCE.[1,6–10] However, ASM OSCs have crucial advantages as compared to polymer OSCs, such as better batch-to-batch reproducibility, high purity, and better processability due to their comparatively higher solubilities.[3,6] The versatility of the available chemical structures results in improved energy level control compared to polymers.[10] Additionally, the strong intermolecular interactions and crystallinity in these systems results in high open circuit voltages and electron mobilities.[7,8] Thus ASM absorber layers represent a promising option for future development of OSC, but their efficiency and stability still need to be improved and better understood.

In order to reach decent performances, OSC absorber layers must form complex structures, the so-called bulk heterojunction (BHJ) morphologies.[11] [12] Thereby, separated donor and acceptor phases form bicontinuous interpenetrating pathways at the nanometer scale. This allows for efficient dissociation of the excitons, which have a very short mean free path in such dielectric materials, and for efficient hole and electron transport in the donor and acceptor phases, respectively.[13–16] Moreover, crystalline phases are desirable to ensure high charge carrier mobilities. The BHJ forms during the fabrication process (coating and further processing steps such as solvent vapor annealing (SVA) or thermal annealing (TA)), so that the processing conditions have a significant impact on the BHJ morphology and thus on the optoelectronic performances.[11] It is therefore interesting to control the morphological features of the BHJ, and for this it is desirable to understand the physical mechanisms driving its formation.

This paper focuses on understanding the morphology formation mechanisms of an ASM photoactive layer upon thermal annealing. The investigated system is a blend of DRCN5T [2,2'-[(3,3''',3'''',4'-tetraoctyl[2,2':5',2'':5'',2''':5''',2''''-quinquethiophene]-5,5''''-diyl)bis[(Z)-methylidyne(3-ethyl-4-oxo-5,2-thiazolidinediylidene)]]bis-propanedinitrile], an oligophiene small molecule donor and $PC_{71}BM$ [[6,6]-Phenyl-$C_{71}$-butyric acid methyl ester], a fullerene acceptor small molecule. Although newer materials allow for better performance, the energy level matching for these molecules still allows the PCE to reach 10%.[4] Most importantly, this system has been studied in details over the past years, giving access to substantial amount of quantitative, precise and usable data on active layer morphology and formation mechanisms. In particular, energy-filtered transmission electron microscopy (EFTEM) investigations allowed us to extract characteristic information on the nanomorphology of the BHJ and analyze the relationship between the post-treatment conditions, the morphology and solar cell performance: the detailed analysis of the microstructure of the DRCN5T crystal fiber lengths and widths for SVA with various solvents, and under different TA temperatures, brought insight into the optimal crystal length leading to the highest PCE.[8,9,17–21] Crystals must be sufficiently large to provide sufficient phase separation and high mobilities, but not too large in order to ensure proper exciton dissociation.[8] However, the process-structure relationship is still not fully unravelled. Therefore, this article aims to shed light on the physical phenomena responsible for the BHJ morphologies observed experimentally in DRCN5T: $PC_{71}BM$ absorber layers, using advanced simulations. Thereby, we focus on the effect of TA post-treatments at different temperatures.[8,9,17–21]

Numerous methods can be used for the simulation of complex phase transformation processes like the ones encountered during the formation of a BHJ. Atomistic, molecular dynamics (MD) or Monte Carlo (MC) simulations can provide insights at the molecular level, but they are time-consuming and restricted to small-scale systems. Continuum mechanics simulation models such as the phase field (PF) approach allow to solve such problems at the mesoscopic level with a reasonable computational effort.[22] PF simulations can provide insight into morphology formation and evolution, into the different phase states, and into the phase transitions during a given process. The intermediate and final states of the morphology can thus be calculated and predicted. The PF approach can describe crystallization,[23–32] including nucleation,[22] isotropic or anisotropic growth, coarsening,[33] and complex solidification structures such as dendrites, spherulites, or eutectic patterns.[22] It can also be used to simulate amorphous-amorphous phase separation (AAPS).[14,23] Since several of these processes may be involved in the morphology formation and evolution of OPV absorber layers, in this paper we propose to use the PF approach to investigate the properties of the DRCN5T: $PC_{71}BM$ blend. All these physical processes can be handled in a PF model recently developed in our group,[23–29,31] so that it can be used to identify the mechanisms driving the BHJ formation of DRCN5T: $PC_{71}BM$ blends upon TA.

The objectives of the present work are (1) to successfully simulate the experimentally observed BHJ morphologies and thus validate the applicability of our PF approach for the investigation of ASM OSC, (2) to study the impact of the different possible physical mechanisms on the BHJ morphology evolution during TA at different temperatures and (3) to identify the physical processes driving the morphology evolution of DRCN5T: $PC_{71}BM$ blends upon TA. In order to do this, we perform PF simulations of the TA process with variable parameters, so as to activate or deactivate the possible physical mechanisms, analyse the consequences on the simulated morphologies and their matching with the measured data.

The remainder of the paper is organized as follows: After an introduction to the method in Sec. 2, we describe in Sec. 3 the results of the simulations, the comparison with the experimental results and the discussions on the drivers of the BHJ formation. Finally, Sec. 4 corresponds to the conclusion and outlook.

# 2. Method

## 2.1. Unravelling BHJ evolution mechanisms with a coupled experimental and simulation approach

In order to unravel the mechanisms driving the BHJ evolution during TA, we first need to determine the different physical phenomena that could potentially be active, the definition of which is shortly given below:

- Crystal nucleation: the spontaneous appearance of new, stable crystals (also called nuclei or germs) from thermal fluctuations, driven by the minimization of the free energy of the system upon transition from the amorphous phase to the crystalline phase.
- Crystal growth: the increase in size of a stable crystal, driven by the minimization of the free energy of the system upon transition from the amorphous phase to the crystalline phase. Thereby, crystals take up material from the amorphous phase.
- Diffusion-limited crystal growth: in this case, diffusion in the amorphous phase is too slow to balance for the material uptake by the growing crystals, so that a depletion zone arises in the amorphous phase around the crystals.
- Crystal (in)stability: Crystal (in)stability is a result of the balance between bulk energy and surface energy of a single crystal. For a crystal to be stable, the bulk energy gain upon growth must be larger than the surface energy increase, otherwise the crystal dissolves. It can be shown that the nuclei have to reach a critical size to be stable and grow.
- Impingement: crystals impinge when they come in contact and form crystal-crystal interfaces, so-called grain boundaries.
- Ostwald ripening: the growth of the largest crystals at the expense of the smallest ones, for crystals separated from each other by the amorphous phase. Material transfer between crystals occurs by transport (diffusion) through the amorphous phase. Ostwald ripening is driven by the minimization of the overall interface energy of the crystal assembly. Thereby, the energy is reduced if smaller crystals shrink and larger crystals grow.
- Grain coarsening: similar to Ostwald ripening, the growth of the largest crystals at the expense of the smallest ones, driven by the minimization of the overall interface energy of the crystal assembly. However, we will use this term for crystals in contact, whereby material transfer between crystals occurs through the grain boundaries.

- Amorphous-amorphous phase separation: the demixing of the amorphous phase driven by donor-acceptor immiscibility. Demixing may occur by spinodal decomposition or nucleation and growth of amorphous demixed regions.

We investigate the significance of these various phenomena during TA of DRCN5T:$PC_{71}BM$ blends by performing simulations with different sets of material parameters, and by then comparing the simulation results with experimental measurements of the morphology. The idea is the following: Despite the outstanding experimental data available on the studied system in the literature (See Sec. 2.2), there is still uncertainty on some model parameters. Depending on these parameter values, the physical phenomena listed above may be active or inactive. Therefore, we vary the undefined parameters on purpose to activate or deactivate independently the physical processes listed above (See Table 2), and then compare the simulated morphologies with experimental data. On the one hand, physical processes inducing a mismatch between the simulated morphology and the experimental observations are concluded to be inactive or negligible during the TA of the DRCN5T:$PC_{71}BM$ film. On the other hand, physical processes responsible for morphologies matching the experimental observations are concluded to be active during the TA of the studied system. At the end, this allows to establish the dominant physical processes driving the morphology evolution, and to finely calibrate the model parameters. Moreover, the comparison of the various simulated morphologies provides a broad general understanding on how crystal nucleation, instability, growth, impingement, grain coarsening, Ostwald ripening, AAPS and diffusion limitations may impact the BHJ morphology.

According to the observations done by EFTEM, the matching between simulations and experiments is evaluated using the following criteria (Ref. [8,21] and Figure 3):

- There are 2 phases at the end of the TA procedure, namely, the crystalline DRCN5T fibers and the amorphous mixed phase. Even if a broad and weak $PC_{71}BM$ peak can be observed by energy filtered electron diffraction measurements, there is no clear evolution of $PC_{71}BM$ aggregation with time and temperature.[8] Further investigations with 4D-scanning confocal electron diffraction measurements showed that the $PC_{71}BM$ phase remains homogeneous.[21] Overall, there is no clear experimental evidence that the $PC_{71}BM$ can crystallize or aggregate during the TA of the system.[8,9,20,21]
- The number of crystals after 10 $min$ of TA decreases with increasing TA temperature.
- The average crystal size increases with increasing TA temperature.
- The crystals conserve their leaf shape upon TA for 10 min.

## 2.2. Simulation procedure and method

***Phase-Field Model.*** The model presented here reduces a more general framework described elsewhere to a binary mixture with one crystallizable material.[24–27,31,32] More specifically, in this work the DRCN5T: $PC_{71}BM$ system is modelled as a mixture where the DRCN5T small molecule (SM) can crystallize, whereas $PC_{71}BM$ remains fully amorphous (See Sec. 2.1). The free energy expression describing such a system can be written as follows.[32]

$$\Delta G_v = \rho\varphi^2\left(g(\phi)W_{fus} + p(\phi)L_{fus}\left(\frac{T}{T_m} - 1\right)\right) + \frac{RT}{v_0}\left(\frac{\varphi\ln\varphi}{N_1} + \frac{(1-\varphi)\ln(1-\varphi)}{N_2} + \varphi(1-\varphi)\chi_{aa} + \phi^2\varphi(1-\varphi)\chi_{ca}\right) + \frac{1}{2}\varepsilon^2(\nabla\phi)^2 + \frac{\pi}{2}{\varepsilon_g}^2|\nabla|\delta(\nabla\theta) + \frac{\kappa(\nabla\varphi)^2}{2} + \Delta G_{num} \tag{1}$$

Here, $\varphi$ and $\phi$ are the volume fraction and crystallization order parameter of the DRCN5T, respectively. The first term of the right-hand side (RHS) of Eq. 1 corresponds to the free energy of crystallization, where $\rho$ is the density of the crystallizable material, $W_{fus}$ the energy barrier that must be overcome upon crystallization, $L_{fus}\left(\frac{T}{T_m} - 1\right)$ the driving force for crystallization, with $L_{fus}$ the heat of fusion, $T$ and $T_m$ the temperature and melting temperature, respectively. The energy barrier has a shape given by the double well function $g = \phi^2(\phi - 1)^2$. An interpolation function $p = \phi^2(3 - 2\phi)$ ensures a smooth transition between the amorphous and the crystalline states. The second term of the RHS corresponds to the free energy of mixing and includes enthalpic and entropic contributions. $R$ is the Boltzmann constant and $v_0$ is the molar volume of the smallest component. The entropic contributions terms are $\frac{\varphi\ln\varphi}{N_1} + \frac{(1-\varphi)\ln(1-\varphi)}{N_2}$ where $N_1$and $N_2$ stand for the molar size of DRCN5T and $PC_{71}BM$, respectively. The enthalpic contributions terms are $\varphi(1-\varphi)\chi_{aa} + \phi^2\varphi(1-\varphi)\chi_{ca}$, where $\chi_{aa}$ is the amorphous-amorphous interaction parameter and $\chi_{ca}$ is the crystalline-amorphous interaction parameter. [28] The third term on the RHS corresponds to the surface energy contributions. $\varepsilon$, $\varepsilon_g$ and $\kappa$ are surface tensions related to the order parameter gradient, to the gradient of the crystal orientation $\theta$ of the crystalline material and to the volume fraction gradients, respectively. Finally, the last term of the RHS $\Delta G_{num} = k(\frac{1}{\varphi} + \frac{1}{1-\varphi})$ is a non-physical term introduced for numerical stability purposes, whereby $k$ is a constant value chosen small enough to avoid any significant impact on the simulated physical phenomena.[12,31,34]

The order parameter and volume fraction time evolution are governed by the stochastic Allen-Cahn (AC) and Cahn-Hilliard (CH) kinetic equations, respectively:

$$\frac{\partial\phi}{\partial t} = -\frac{N_1 v_0}{RT}M_{aniso}(\theta)\left(\frac{\partial\Delta G_v}{\partial\phi} - \nabla\frac{\partial\Delta G_v}{\partial\nabla\phi}\right) + \zeta_{AC} \tag{2}$$

$$\frac{\partial\varphi}{\partial t} = \frac{v_0}{RT}\nabla\left[\lambda(\varphi,\phi)\nabla\left(\frac{\partial\Delta G_v}{\partial\phi} - \nabla\frac{\partial\Delta G_v}{\partial\nabla\phi}\right)\right] + \zeta_{CH} \tag{3}$$

Here, $M_{aniso}$, $\lambda$, $\zeta_{AC}$ and $\zeta_{CH}$ are the AC mobility, Onsager mobility, the AC and CH fluctuations on the order parameter and the volume fraction respectively. The AC kinetics controls the crystallization, whereas the CH kinetics controls the material transport.

The growth of DRCN5T fibers is anisotropic, thus requiring an orientation-dependent implementation of at least one relevant growth parameter. Anisotropic crystallization of DRCN5T is due to a preferential spatial arrangement of the molecules in one direction which cannot be accounted for through continuum mechanics models such as the PF. However crystals anisotropy can be modelled with the PF phenomenologically.[35] Commonly, the anisotropy is included through the AC mobility $M_{aniso}$ and/or the surface tension parameter $\varepsilon$.[36,37] In this work, the anisotropy was implemented on the AC mobility $M_{aniso}$ for simplicity, which is given by

$$M_{aniso}(\theta) = M_0 \frac{1}{1+M_{ratio}}(1 + M_{ratio} * f_\theta(\theta)). \quad (4)$$

In this equation, $M_0$ is a constant coefficient fixing the time scale for crystallization,[31] and $M_{ratio}$ determines the longitudinal to transversal growth rate ratio of the DRCN5T crystals $M_{ratio} + 1$. $f_\theta$ is a sinusoidal anisotropy function (see Sec. 1 in SI), which depends on the angle $\theta$ between the normal to the crystal interface and the crystal's first principal axis in the following way:

$$f_\theta(\theta) = \frac{cos(2\theta) + 1}{2} \quad (5)$$

If the normal to the interface is aligned with the crystal's first principal axis, $f_\theta = 1$ and $M_{aniso} = M_0$. If the normal to the interface is perpendicular to the crystal's first principal axis, $f_\theta = 0$ and $M_{aniso}(\varphi, \phi) = M_0 \frac{1}{1+M_{ratio}}$. Other anisotropy functions have been evaluated, but the function above models the anisotropic behavior of DRCN5T crystallization in the DRCN5T:$PC_{71}BM$ blend observed in the EFTEM measurements most accurately based on the criteria of conservation of the principal-to-minor axis ratio during growth and the reproduction of the anisotropic leaf-like shape observed experimentally (see Sec. 1 in SI).

Finally, we can choose between the slow and fast mode dependencies for the Onsager mobility $\lambda$ for the self-diffusion coefficients.[38,39] This work uses the slow mode theory so that the slowest component controls diffusion. The expression of the Onsager mobility is then $\lambda = \omega_1(1 - \frac{\omega_1}{\omega_1+\omega_2})$, where $\omega_1 = N_1\varphi D_{s,1}(\varphi, \phi)$ and $\omega_2 = N_2(1-\varphi)D_{s,2}(\varphi, \phi)$. The self-diffusion coefficients of DRCN5T and $PC_{71}BM$ are calculated using the Vignes' Law, $D_{s,1}(\varphi, \phi) = t(\phi)({D_{s,1}}^{amorph^{\varphi\to 1}})^{\varphi}({D_{s,1}}^{amorph^{\varphi\to 0}})^{(1-\varphi)}$ and $D_{s,2}(\varphi, \phi) = t(\phi)({D_{s,2}}^{amorph^{\varphi\to 0}})^{\varphi}({D_{s,2}}^{amorph^{\varphi\to 1}})^{(1-\varphi)}$, respectively. Thereby, ${D_{s,1}}^{amorph^{\varphi\to 1}}$, ${D_{s,1}}^{amorph^{\varphi\to 0}}$, ${D_{s,2}}^{amorph^{\varphi\to 0}}$ and ${D_{s,2}}^{amorph^{\varphi\to 1}}$ are the self-diffusion coefficients of DRCN5T in its pure amorphous phase, of DRCN5T in a pure amorphous $PC_{71}BM$ phase, of $PC_{71}BM$ in its pure amorphous phase, and of $PC_{71}BM$ in a pure amorphous DRCN5T phase. $t$ is a penalty function that decreases the value of the self-diffusion coefficient when the crystallinity $\phi$ increases.[31]

***<u>Simulation setup.</u>*** The starting point of the simulation is the morphology of the as-cast film.[8,21] The volume fraction of DRCN5T in the investigated blend is $\varphi = 0.62$.[8] The blend consists of two phases: The leaf-shaped DRCN5T crystals are embedded in the second phase, an amorphous mixed region of DRCN5T and $PC_{71}BM$. The DRCN5T crystals are assumed to be very pure, i.e., their equilibrium volume fraction is $\varphi \approx 1$. As an initial condition, 120 elliptic nuclei following a Gaussian size distribution between $l * w = 17 * 7\ nm^2$ and $l * w = 27\text{x}11\ nm^2$ (where $l$ and $w$ are the crystal length and width, respectively) are randomly placed in the 2D simulation box.[40] The box has a size of 512*512 $nm^2$, similar to the films measured by EFTEM, with periodic boundary conditions in both directions and a resolution of 1 nm. The equations are solved using a finite volume approach. Regarding the time stepping scheme, Pareschi Russo's second order diagonally implicit Runge-Kutta method was chosen because it provides the optimal compromise between computation time and accuracy. The simulated TA temperatures are 80°$C$, 100°$C$, 120°$C$, 140°$C$, and 160°$C$.

***Parameter determination.*** As far as possible, the simulation parameters were extracted from the literature. The smallest component of the studied system is the $PC_{71}BM$, with a calculated molar volume $v_0 = 5.69 * 10^{-4}\ m^3/mol$ and calculated molecular sizes of $N_1 = 1$ and $N_2 = 1.75$.[41–44] The enthalpy of crystallization for pure DRCN5T is $L_{fus} = 44{,}89\ kJ\ kg^{-1}$ and its melting temperature $T_m = 490.25\ K$.[43,44] Since the temperature varies between the different TA experiments conducted,[8] the temperature dependencies of the surface tension parameter $\varepsilon$, the amorphous-amorphous interaction parameter $\chi_{aa}$, the crystalline-amorphous interaction parameter $\chi_{ca}$, the self-diffusion coefficients $D_{ij}$, and the energy barrier $W_{fus}$ are taken into account as summarized in Table 1.[22,29,45,46] The temperature dependency of the self-diffusion coefficients $D_{ij}$ (Table 1) of a material $i$ in an amorphous matrix of a pure material $j$ is derived from the Stokes-Einstein (SE) equation for a spherical particle,[45] $D_{ij} = \frac{k_B T}{6\pi\eta(T) r_p}$, where $k_B$ is the Boltzmann constant, and $r_p$ is the radius of a spherical particle moving in an amorphous matrix of viscosity $\eta$. The matrix viscosity $\eta$ follows an Arrhenius law $\eta(T) \propto e^{\frac{\beta_D}{T}}$,[45] which leads to $D_{ij} = \alpha_D T e^{-\frac{\beta_D}{T}}$, where $\alpha_D$ and $\beta_D$ are constants. The mobility $M_{aniso}$ is assumed to follow a similar Arrhenius law because crystal growth results from the mobility of the species attaching to the crystal interface.[47] The temperature dependency is attributed through the coefficient $M_0 = \alpha_{M_0} T e^{-\frac{\beta_D}{T}}$, where $\alpha_{M_0}$ is a constant coefficient. The amorphous-amorphous interaction parameter $\chi_{aa}$ follows the empirical law $\chi_{aa} = \alpha_{\chi_{aa}} + \frac{\beta_{\chi_{aa}}}{T}$ where $\alpha_{\chi_{aa}}$ and $\beta_{\chi_{aa}}$ are constant coefficients. [46] $\alpha_{\chi_{aa}}$ and $\frac{\beta_{\chi_{aa}}}{T}$ represent the entropic and enthalpic part of the interaction parameter, respectively. The crystalline-amorphous interaction parameter $\chi_{ca}$ is chosen to be inversely proportional to the temperature, $\chi_{ca} = \alpha_{\chi_{ca}} \frac{L_{fus}}{RT}$, as proposed by Matkar and Kyu,[29] where $\alpha_{\chi_{ca}}$ is a constant. Finally, the surface tension coefficient is assumed to evolve as $\varepsilon = \alpha_\varepsilon \sqrt{T}$ and the energy barrier as $W_{fus} = \alpha_W T$ as described in Table 1. As a result, the crystal interface thickness $\delta \propto \frac{\varepsilon}{\sqrt{W_{fus}}}$ is constant and the surface energy $\sigma \propto \varepsilon\sqrt{W_{fus}} \propto \sqrt{T}$ increases linearly with temperature. The temperature dependencies of $\varepsilon$, $W_{fus}$, $\sigma$ and the fact that the interface thickness does not evolve with temperature is consistent with molecular dynamics simulations.[22,30,48,49] $\alpha_\varepsilon$ and $\alpha_W$ are constant coefficients. Table 1 summarizes the temperature dependency of the parameters.

| **Parameter** | **Temperature dependency** |
|---|---|
| **Self-diffusion coefficients** | $\boldsymbol{D_{ij} = \alpha_D T e^{-\frac{\beta_D}{T}}}$ [Ref[45]] |
| **Mobility coefficient** | $\boldsymbol{M_0 = \alpha_{M_0} T e^{-\frac{\beta_D}{T}}}$ |
| **Amorphous-amorphous interaction parameter** | $\boldsymbol{\chi_{aa} = \alpha_{\chi_{aa}} + \frac{\beta_{\chi_{aa}}}{T}}$ [Ref [46]] |
| **Amorphous-crystalline interaction parameter** | $\boldsymbol{\chi_{ca} = \alpha_{\chi_{ca}} \frac{L_{fus}}{RT}}$ [Ref [29]] |
| **Surface tension related to the order parameter gradients** | $\boldsymbol{\varepsilon = \alpha_\varepsilon \sqrt{T}}$ [Ref [22]] |

| Energy barrier upon crystallization | $W_{fus} = \alpha_W T$ [Ref [22]] |
|---|---|

*Table 1: Temperature dependency of materials parameters*

This lets us with the parameters $\alpha_D$, $\beta_D$, $\alpha_{M_0}$, $\alpha_{\chi_{aa}}$, $\beta_{\chi_{aa}}$, $\alpha_{\chi_{ca}}$, $\alpha_\varepsilon$ and $\alpha_W$ which can be adjusted to match available experimental data. First of all, for a given material parameter set, the phase diagram of the mixture is calculated and the interaction parameters $\alpha_{\chi_{aa}}$, $\alpha_{\chi_{ca}}$ and $\beta_{\chi_{aa}}$ are tuned so that the resulting phase diagram (1) matches the experimental liquidus points of the DRCN5T:$PC_{71}$BM phase diagram,[43] and (2) shows a solidus curve with volume fractions close to $\varphi \approx 1$ (pure DRCN5T crystals) at all temperatures. The simulated phase diagrams for a miscible blend (simulations type A, B, C, and E) and for an immiscible blend (simulations type D) are shown in Sec. 2 of the SI. In both cases, the dependence of the melting temperature to volume fraction (melting point depression) is recovered (See Sec. 2 of SI). Moreover, the parameters $\alpha_\varepsilon$ and $\alpha_W$ are chosen to meet two requirements. On the one hand, the interface thickness $\delta \propto \frac{\varepsilon}{\sqrt{W_{fus}}}$ spreads over 8 mesh points for proper numerical convergence.[30] On the other hand, the critical radius is adjusted such that nuclei become unstable at high annealing temperatures: in the current PF simulations, the critical radius is estimated from the classical nucleation theory (CNT). It can be expressed in terms of phase field parameters for 2D simulations of pure isotropic materials as $r^* = \frac{\varepsilon\sqrt{W_{fus}}}{3L_{fus}(1-\frac{T}{T_m})\sqrt{2\rho}}$ (See Sec. 3 in SI) and is a first indicator for the stability of the ellipsoidal crystals. Indeed, the local radius of curvature should be greater than the critical radius of curvature $r^*$ for the crystal interface to propagate. Although the critical radius expression is different in a blend and should, in any case, depend on the volume fraction, this expression is a reasonable first approximation. It indicates a temperature dependency $r^* \propto \frac{T}{(1-\frac{T}{T_m})}$. The coefficients $\alpha_{M_0}$ and $\beta_D$ are chosen such that the simulated DRCN5T crystal growth rates match the experimentally measured growth rates, which have been obtained from the time dependence of the crystal widths and lengths measured at the different annealing temperatures. The growths rates are constant over time, as expected for simple growth (See Sec. 4 of the SI).[8] Indeed, there should be a match between the experimental and the simulated longitudinal and transversal growth rates of the DRCN5T crystal fibers. Therefore, the mobility coefficient $M_0$ and the ratio $M_{ratio}$ are adjusted to recover the longitudinal and transversal growth rates measured by ex-situ measurements.[8] The full set of parameters for the different simulations is given in Sec. 5 of the SI.

***Activating various sets of physical mechanisms.*** Following the method described in Sec. 2.1, five types of simulations are performed to elucidate the physical phenomena occurring during the TA of the investigated system. Each type of simulation corresponds to a set of parameters chosen to activate different physical processes, as summarized in Table 2. Simulations of type A correspond to the “reference” simulations, whereby the interplay between three different phenomena conditions the morphology evolution: DRCN5T crystal growth, Ostwald ripening, and the temperature-dependent stability of the nucleus (see Sec. 3). The initial nucleus density is sufficiently low to prevent impingement during the TA time. In addition, the self-diffusion coefficients are high enough to prevent diffusion-limited crystal growth. Furthermore, the amorphous-amorphous interaction parameter is below the critical value to inhibit AAPS. Thermal fluctuations are not activated, which disables nucleation and grain coarsening. In simulations of type B, the initial crystal density is increased to cause crystal impingement during the simulated TA time. This allows us to check the impact of impingement on the evolution of morphology. In simulations of type C, thermal fluctuations are added to activate nucleation and grain coarsening, in order to investigate their impact on the morphology evolution. Note that we distinguish here between “grain coarsening” and “Ostwald ripening”. Both processes relate to the growth of large crystals at the expense of smaller ones due to surface tension forces. However, as stated above and following the definition used in metallurgy, we define grain coarsening as the evolution of grains in contact with each other involving material transfer at grain boundaries, whereas the term “Ostwald ripening” will be used to describe coarsening for grains that are not in contact, involving material transfer through the amorphous phase. In simulations of type D, the amorphous-amorphous interaction parameter is set above its critical value to trigger AAPS and characterize the resulting morphology. Finally, the impact of diffusion-limited crystal growth is studied in simulations of type E. In this last case, due to the very low values chosen for the self-diffusion coefficients, the rate of the amorphous DRCN5T material migration towards the crystals is slower than the rate of DRCN5T attachment to the crystal interface.

| Simulation Type / Active Physical Phenomena | A | B | C | D | E |
|---|---|---|---|---|---|
| Crystal Growth, Ostwald Ripening, and Nucleus Stability | YES | YES | YES | YES | YES |
| Diffusion Limitation | NO | NO | NO | NO | YES |
| Crystals Impingement | NO | YES | YES | NO | NO |
| AAPS | NO | NO | NO | YES | NO |
| Nucleation and Grain Coarsening | NO | NO | YES | NO | NO |

*Table 2: Simulated physical phenomena for the TA of an all-small molecules bulk heterojunction.*

# 3. Results and Discussions

## 3.1. Crystal growth, stability and Ostwald ripening

In simulations of type A, only crystal growth, Ostwald ripening, and nucleus stability can play a role. The objective is to determine whether this set of active processes can lead to successful replication of the experimental observations detailed in Sec. 2.1. This will allow to conclude whether crystal growth, Ostwald ripening, and nucleus stability are significantly active during the thermal annealing of DRCN5T:$PC_{71}BM$ active layers. The morphology evolution for a TA temperature of 140 $°C$ is shown in Figure 2 at annealing times of 0 $s$, 150 $s$, 300 $s$, 450 $s$, 600 $s$, and 900 $s$ (Figure 2a). The DRCN5T volume fraction in the crystals is $\varphi \approx 1$ as expected from the DRCN5T:$PC_{71}BM$ solidus line (See Sec. 2 in SI). Several small crystals dissolve before 300 $s$ of TA (Figure 2b-c). Starting from 300 $s$ of TA (Figure 2c), the remaining, stable DRCN5T crystals grow in size, by taking up material from the amorphous phase, as long as the amount of DRCN5T in the amorphous phase is above its equilibrium value. This leads to a progressive and significant decrease of the DRCN5T volume fraction in the amorphous phase until a TA time of 900 s (Figure 2f). The growth is anisotropic, and the leaf shape of the crystals is conserved for TA times below 600 $s$ (Figure 2b-e). However, at TA times beyond 600 $s$, the DRCN5T crystals impinge (Figure 1f), and their shape is no longer conserved (see Sec. 3.2).

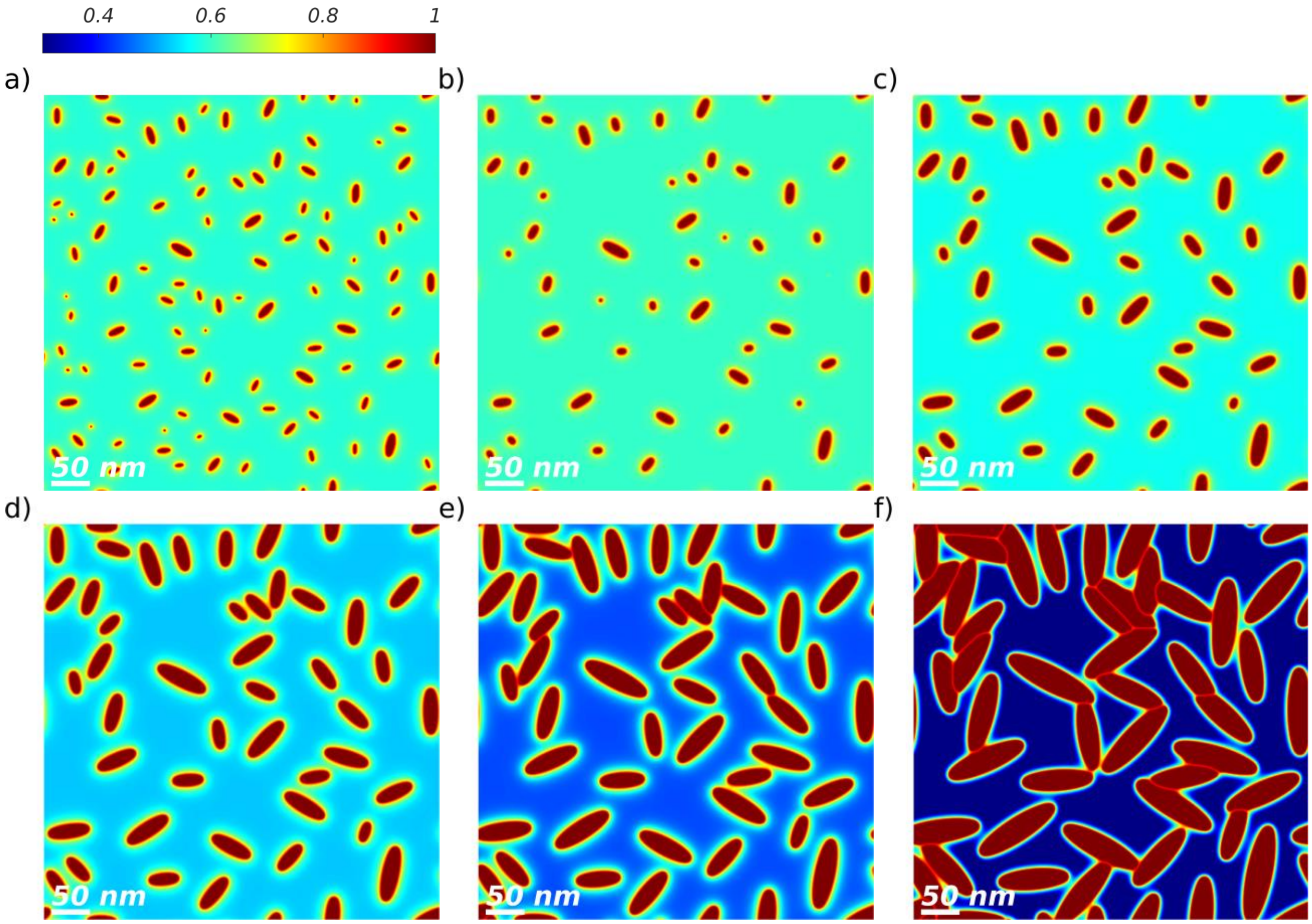

*Figure 2: TA at 140°C of a binary DRCN5T:$PC_{71}BM$ blend with initial blend ratio 1:0.8, simulation of type A. The DRCN5T volume fraction fields are shown for the as-cast film with low crystal density (a) and after TA of 150 $s$ (b), 300 $s$ (c), 450 $s$ (d), 600 $s$ (e), and 900 $s$ (f). The volume fraction colorbar spreads from dark blue (Low DRCN5T content) to dark red (pure DRCN5T).*

Simulations of type A are performed at temperatures of 80 $°C$, 100 $°C$, 120 $°C$, 140 $°C$ and 160 $°C$, and the simulated morphologies after 10 $min$ of TA are compared to the EFTEM images in Figure 3. Even though crystal overlap and the growth of face-on crystals observed in the EFTEM samples are not taken into account in the 2D simulations, the theoretical results nicely match the experimental measurements according to the criteria listed in Sec. 2.1, namely the presence of 2 phases, the decrease of the number of DRCN5T crystals, the increase of DRCN5T crystal sizes and the conservation of the leaf shape. At 80 $°C$, crystal growth is very limited (Figure 3b). On the contrary, crystal growth is significant at TA temperatures of 100 $°C$, 120 $°C$, 140 $°C$, and 160 $°C$ (Figure 3c-f), and the final crystal size increases with increasing temperature. This is due to the nearly exponential temperature dependency of the crystal growth rate (See Sec. 2).

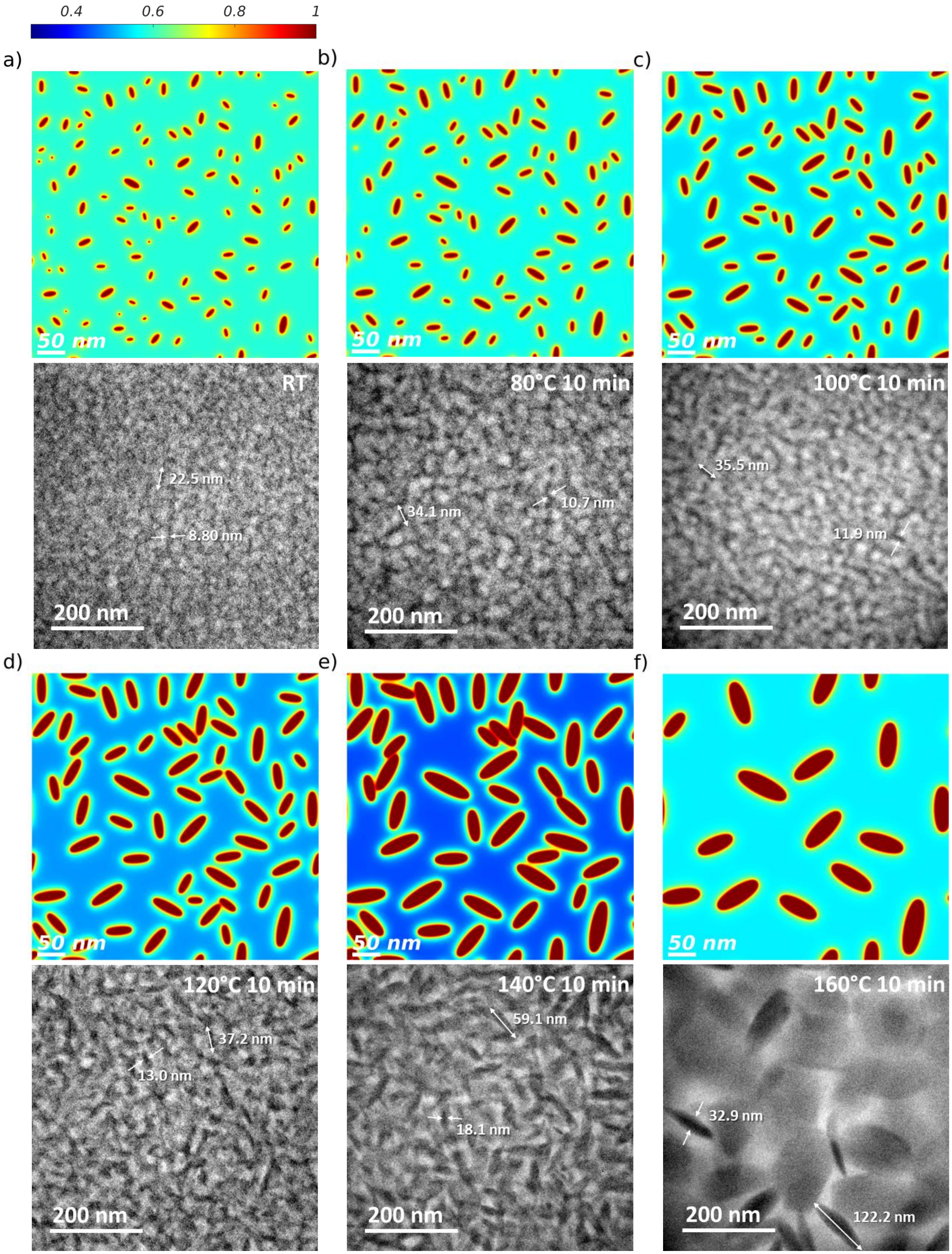

*Figure 3: TA of a binary DRCN5T:PC$_{71}$BM blend with initial blend ratio 1:0.8, simulation of type A. The DRCN5T volume fraction fields versus EFTEM images (Reproduced with permission from*

*Ref.21 CC BY 4.0) measured are shown for the as-cast film with low crystals density (a) and after 600 $s$ at TA temperatures of 80 $°C$ (b), 100 $°C$ (c), 120 $°C$ (d), 140 $°C$ (e), and 160 $°C$ (f). The volume fraction colorbar spreads from dark blue (Low DRCN5T content) to dark red (pure DRCN5T). The EFTEM images are elemental maps of carbon, the grayscale of the phase's density spreads from light grey (carbon-rich phase, $PC_{71}BM$) to dark (carbon-poor phase, DRCN5T). The large platelets are assigned to face-on crystals (not considered in the simulations) whereas the more pronounced fibers correspond to edge-on crystals.*

Moreover, like in the experimental observations, the higher the temperature, the fewer crystals are observed in the final morphology (see also Figure 4). This is due to the increasing critical radius with temperature (See Sec. 2). At all TA temperatures, it has been checked that crystals dissolve whenever they are smaller than the critical size (See Sec. 6 in SI). In addition, note that the volume fraction of DRCN5T in the amorphous phase after 10 $min$ of TA varies with temperature, decreasing from 80 $°C$ to 140 $°C$, and increasing again for 160 $°C$. This is a purely kinetic effect: crystallization is not complete after $10\,min$ and thus the equilibrium volume fraction of the amorphous phase (which is fairly constant in the range 80-160 $°C$, see Figure S4) is not reached yet. The time-dependent DRCN5T amount in the amorphous phase is the result of the (nearly negligible) initial DRCN5T release from instable germ dissolution, and mainly of DRCN5T consumption for crystal growth. DRCN5T consumption is faster for higher temperatures, but the number of stable crystals taking up materials decreases with temperature. The balance of both results in a maximum uptake from the amorphous phase at intermediate temperatures (140 $°C$, see Figure 4).

Note that the DRCN5T volume fraction decrease in the liquid phase is additionally responsible for an increase of the critical radius with time. However, we verified that this volume fraction decrease does not result in further crystal dissolution, as shown in Figure 4. The change in the number of crystals and the change of the DRCN5T volume fraction in the amorphous phase $\varphi_{am}$ are actually uncorrelated. At 120 $°C$ for instance, the crystals dissolve until *200 s* of TA due to the initial temperature rise, whereby the DRCN5T volume fraction is constant during this phase. Afterwards, the DRCN5T volume fraction decreases, which indeed leads to an increase of the critical size. However, the remaining crystals grow sufficiently fast for their size always remaining larger than the critical size, and therefore the number of crystals remains constant. Such a behaviour is observed for all other TA temperatures.

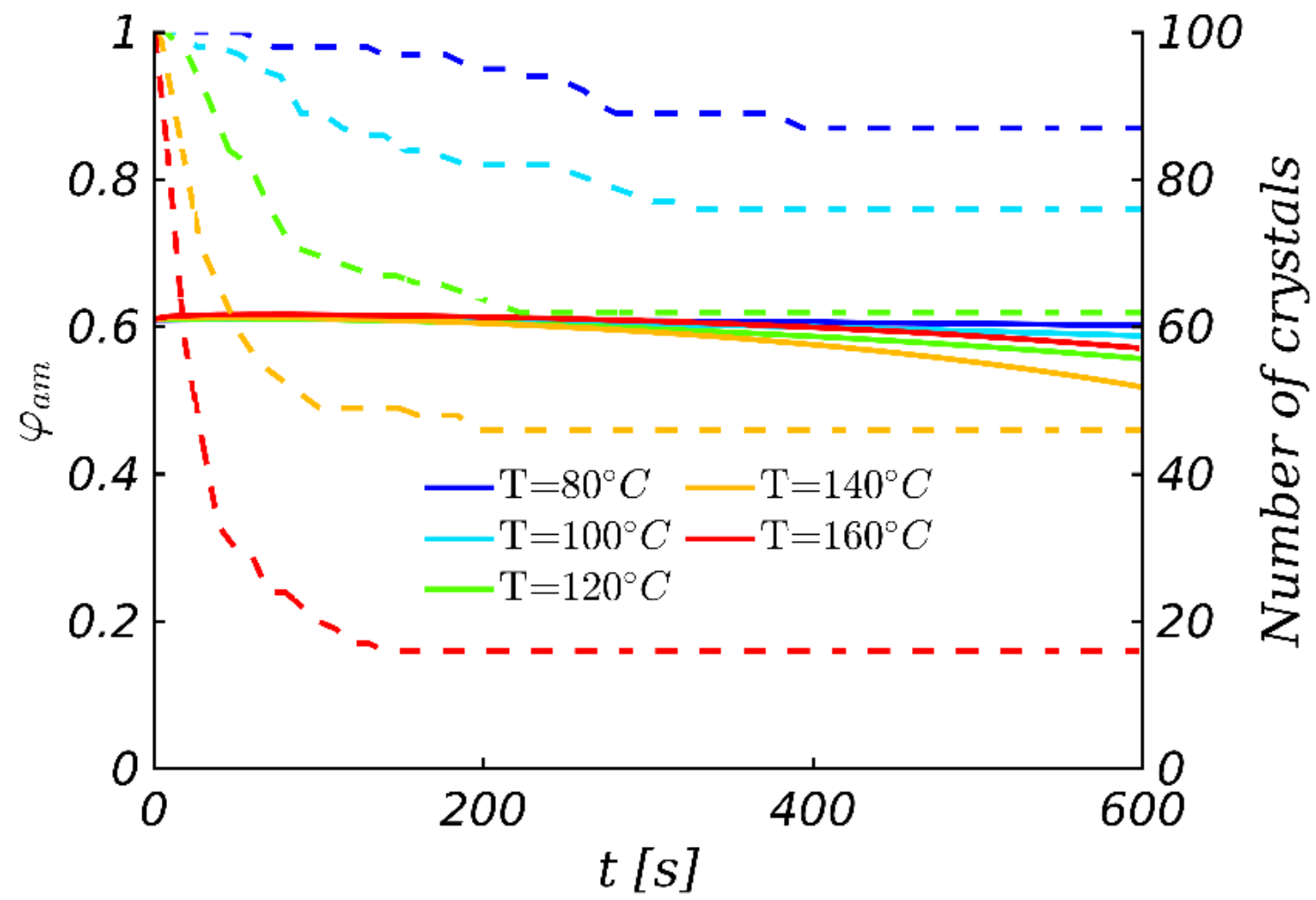

*Figure 4: Time-dependent DRCN5T volume fraction in the amorphous phase $\varphi_{am}$ (solid line and left y-axis) and number of DRCN5T crystals (dashed line and right y-axis) for different annealing temperatures.*

In the simulations presented above, the evolution towards a smaller number of larger crystals upon TA is due to initial dissolution of the smaller crystals due to thermal instability, and growth of the remaining, stable crystals. Ostwald ripening, which is by definition a process leading to disappearing of smaller crystals and size increase of the larger ones, could also be invoked to justify the experimental morphologies. However, in the simulations, it is negligible. This is because Ostwald ripening becomes dominant only at late stages of the crystallization process. The evolution of the morphology is driven by the free energy minimization of the system. This actually starts with the minimization of the bulk energy due to crystal growth. Ostwald ripening, which is driven by the reduction of crystal interfacial energy, only becomes dominant once the bulk energy does nearly not change anymore (equilibrium compositions are nearly reached in the amorphous and crystalline phases). However, the equilibrium volume fractions are far from being reached at the end of the annealing (see Figure 4), so that the growth process fully masks the consequences of Ostwald ripening.

Nevertheless, these findings assume that the DRCN5T volume fraction in the amorphous phase of the as-cast film, at the beginning of the TA, is far above the equilibrium value. However, it is actually unknown experimentally. If one assumes an initial DRCN5T volume fraction equal or close to the equilibrium volume fraction, crystal growth would be only shortly active and Ostwald ripening would become quickly dominant. However, in such conditions, the average crystal size would follow a time evolution proportional to the cubic root of time or even slower,[23,50–52] whereas it increases linearly with time in the case of crystal growth (see Figure 5). Experimental evaluation showed that in the case of TA, the average DRCN5T crystal size evolves linearly with time.[8] This finding supports the idea that Ostwald ripening is not significant during TA of DRCN5T:$PC_{71}$BM.

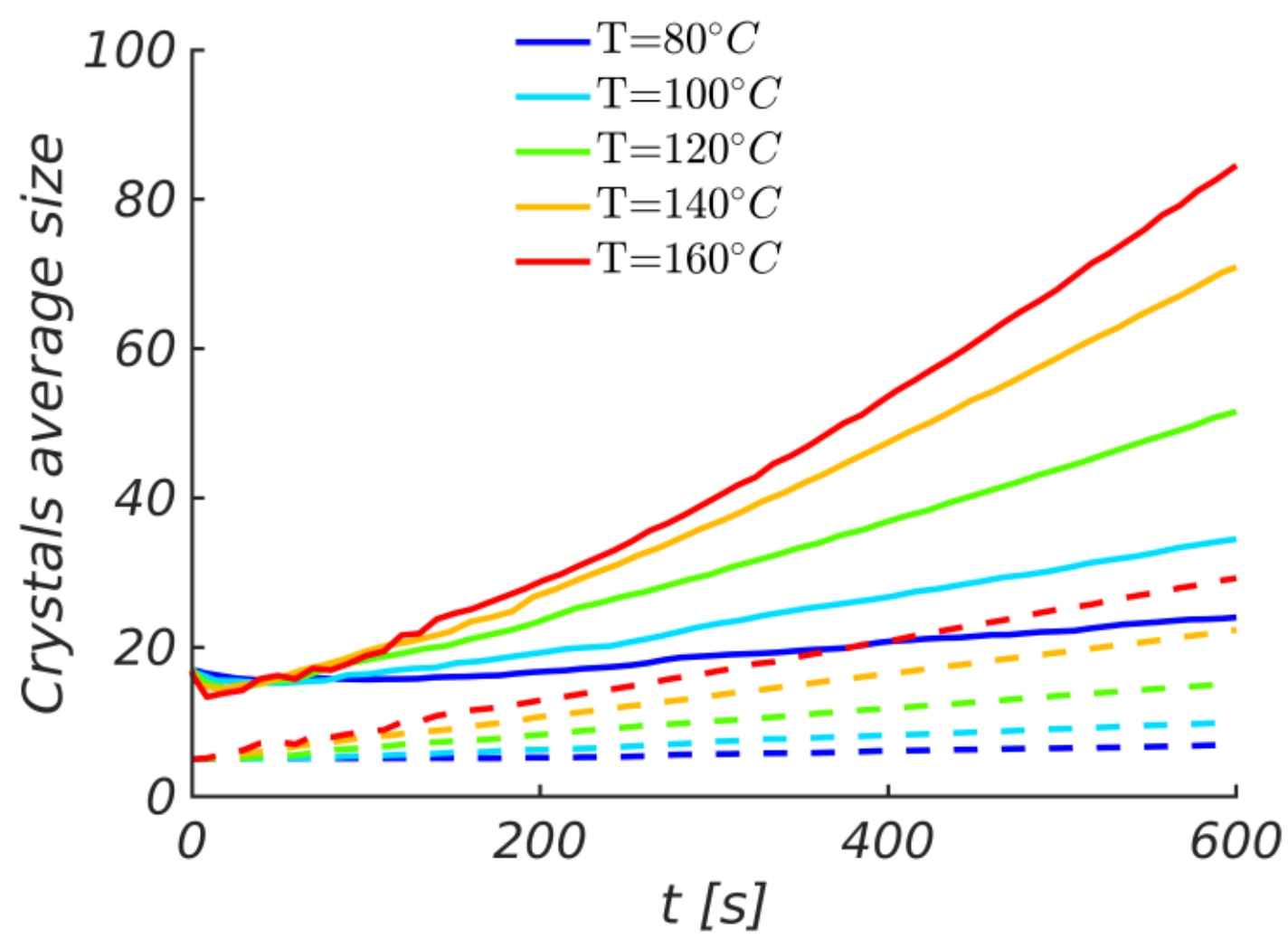

*Figure 5: Time evolution of the average crystal lengths (solid lines) and widths (dashed lines) for different annealing temperatures.*

Overall, the matching between morphologies resulting from simulations of type A and the experimental observations allows us to conclude that the morphology evolution of the as-cast DRCN5T:$PC_{71}BM$ blends during TA is mainly driven by the dissolution of initially too small, unstable germs and the subsequent growth of the remaining large, stable DRCN5T crystals.

## 3.2. Crystal impingement

As shown in Figure 3b-f, at 10 $min$ of TA, the leaf shape of the crystals is conserved at all annealing temperatures. This is because the density of growing crystals is too low for the crystals to impinge within 10 $min$ of TA. However, the crystal density of the as-cast film was not determined experimentally.[8] In simulations of type B, the initial crystals density is increased, which increases the chance for crystals to impinge. The objective is to determine whether the presence of impingement can still lead to BHJ morphologies similar to the experimental observations of Sec. 2.1. This will allow to conclude whether crystals impingement is significantly active during the thermal annealing of DRCN5T:$PC_{71}BM$ active layer. Wherever impingement is happening, crystal growth is hindered at the crystal-crystal boundaries. As compared to simulations of type A, this results in a significant deviation of the crystal shape from the leaf shape for the impinged crystals at the contact points where growth is hindered, as shown in Figure 6d-f for TA temperatures above 120 $°C$. This is not in line with the experimentally observed morphologies, which support the idea that the density of growing crystals during TA is too low for the crystals to impinge significantly. The mismatch between morphologies resulting from simulations of type B and the experimental observations allows us to conclude that crystals impingement is not significantly active during the 10 min of TA of the DRCN5T: $PC_{71}BM$ active layer.

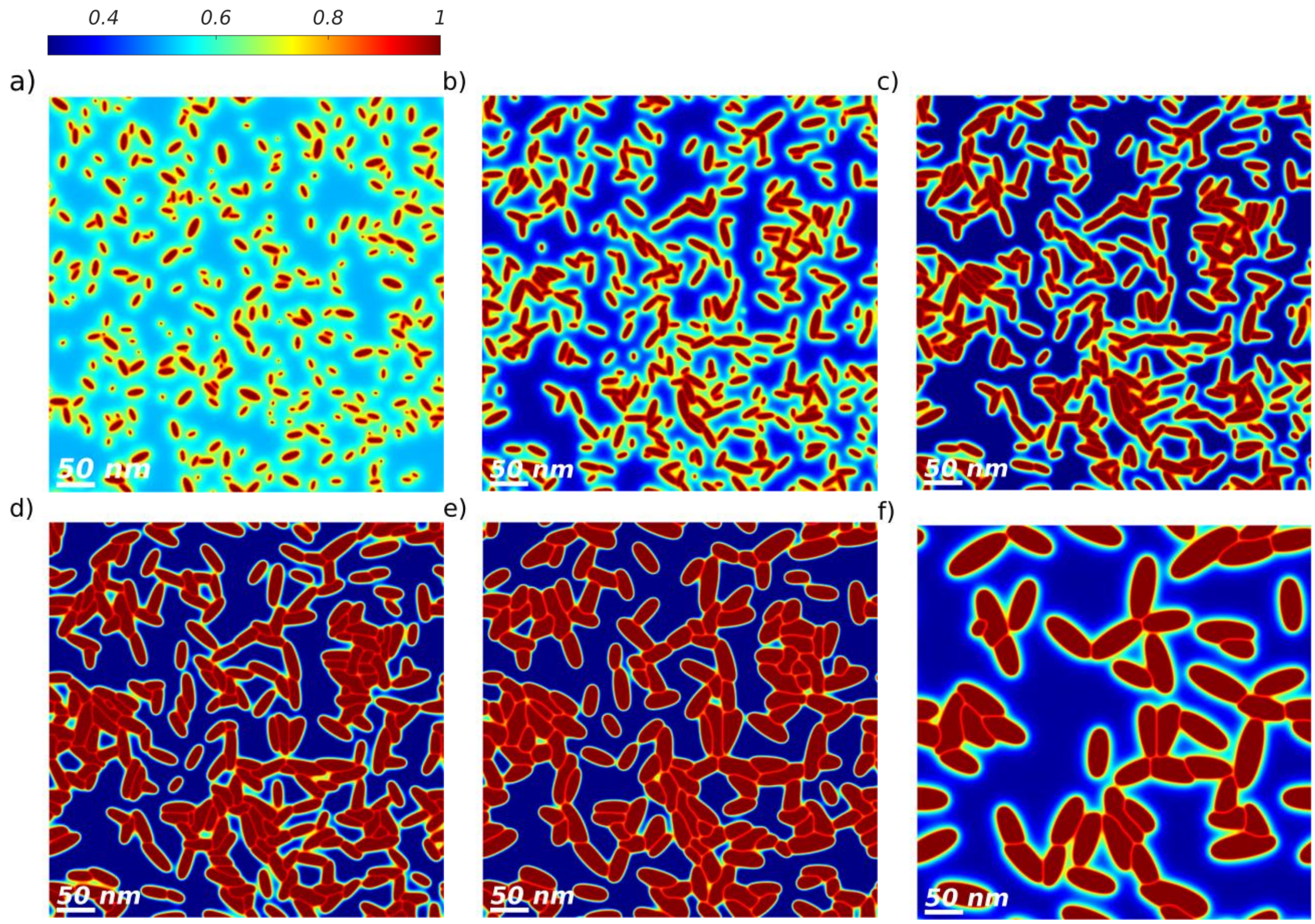

*Figure 6: TA of a binary DRCN5T:$PC_{71}BM$ blend with initial blend ratio 1:0.8, simulation of type B. The DRCN5T volume fraction fields are shown for the as-cast film with high crystal density (a) and after 600 $s$ at TA temperatures of 80 $°C$ (b), 100 $°C$ (c), 120 $°C$ (d), 140 $°C$ (e), and 160 $°C$ (f)). The volume fraction colorbar spreads from dark blue (Low DRCN5T content) to dark red (pure DRCN5T).*

## 3.3. Nucleation and grain coarsening

The effect of nucleation and grain coarsening (i.e. the growth of largest crystals at the expense of the smallest ones driven by the minimization of grain boundary energy) is investigated through simulations of type C, whereby thermal fluctuations are activated. The comparison of the experimental observations of Sec. 2.1 with morphologies influenced by nucleation and coarsening will allow to conclude whether both processes are significantly active during the thermal annealing of DRCN5T:$PC_{71}BM$ active layer. First of all, the morphology evolution at the lowest annealing temperature (80 $°C$) is investigated (Figure 7, top row), for which the largest nucleation rate (See Sec. 3 in SI) and the lowest growth rate (See Sec. 5 in SI) are expected. The morphology is very different as compared to simulations of type A. On the one hand, the number of crystals after 150 $s$ to 600 $s$ of TA (Figure 7, top row) is considerably increased due to nucleation, unlike the observations from the experiments.[8,21] The growth of these many crystals leads to a quick decrease of the DRCN5T content in the amorphous phase. On the other hand, the nucleation process favours the development of morphologies with clustered crystals and large-scale phase separation, which can be explained in three main steps. The first step is the favoured DRCN5T germ nucleation at the existing crystals interface where the DRCN5T volume fraction is high (Figure 7a-b). The second step is the formation of amorphous phase bridges with high DRCN5T content between the germs favoured by the limitation of the energetic cost associated with amorphous-crystalline interfaces. The third step is the further nucleation within the formed amorphous bridges, because the critical germ size is smaller and the driving force of crystallization is larger for higher DRCN5T content.[27] This large scale co-continuous morphology favours crystal impingement. Here again, this leads to ill-defined crystal shapes, as observed in Figure 7c. At higher temperatures (for example 140$°C$, Figure 7, bottom row), nucleation is less intense, and growth is faster, so that the evolution is more growth dominated. There are less but larger crystals growing in the BHJ than at lower temperatures. As a result, the observations made at 80 $°C$ regarding the number of crystals, crystal impingement and clustering, and large-scale phase separation are less pronounced, but they are still valid. Thus, the nucleation process generates morphological features that are not observed experimentally, and we conclude that nucleation is not significantly active during TA of the studied system.

Similar to Ostwald ripening, grain coarsening becomes dominant when the equilibrium volume fraction is reached in the amorphous phase. This occurs at very late stages (note that at both annealing temperatures the DRCN5T volume fraction in the amorphous phase still decreases with time until 600 $s$) and explains why it is a secondary process during the 10 $min$ of TA of the DRCN5T: $PC_{71}BM$ film. The influence of coarsening can be evaluated by counting the number of stable crystals that disappear upon coarsening. At 140 $°C$ for instance, it is observed that crystal dissolution due to coarsening only occurs starting from 250 $s$ of TA (See Sec. 7 in SI). At the end of the TA (600 $s$), only 15-20% of the crystals have disappeared due to coarsening. The impact of the coarsening is, therefore, very limited overall and cannot justify the strong reduction of the number of crystals experimentally observed at high TA temperatures.

Overall, the mismatch between morphologies resulting from simulations of type C and the experimental observations allow us to conclude that nucleation and grain coarsening are not significantly active during the 10 $min$ of TA of the DRCN5T: $PC_{71}BM$ active layer.

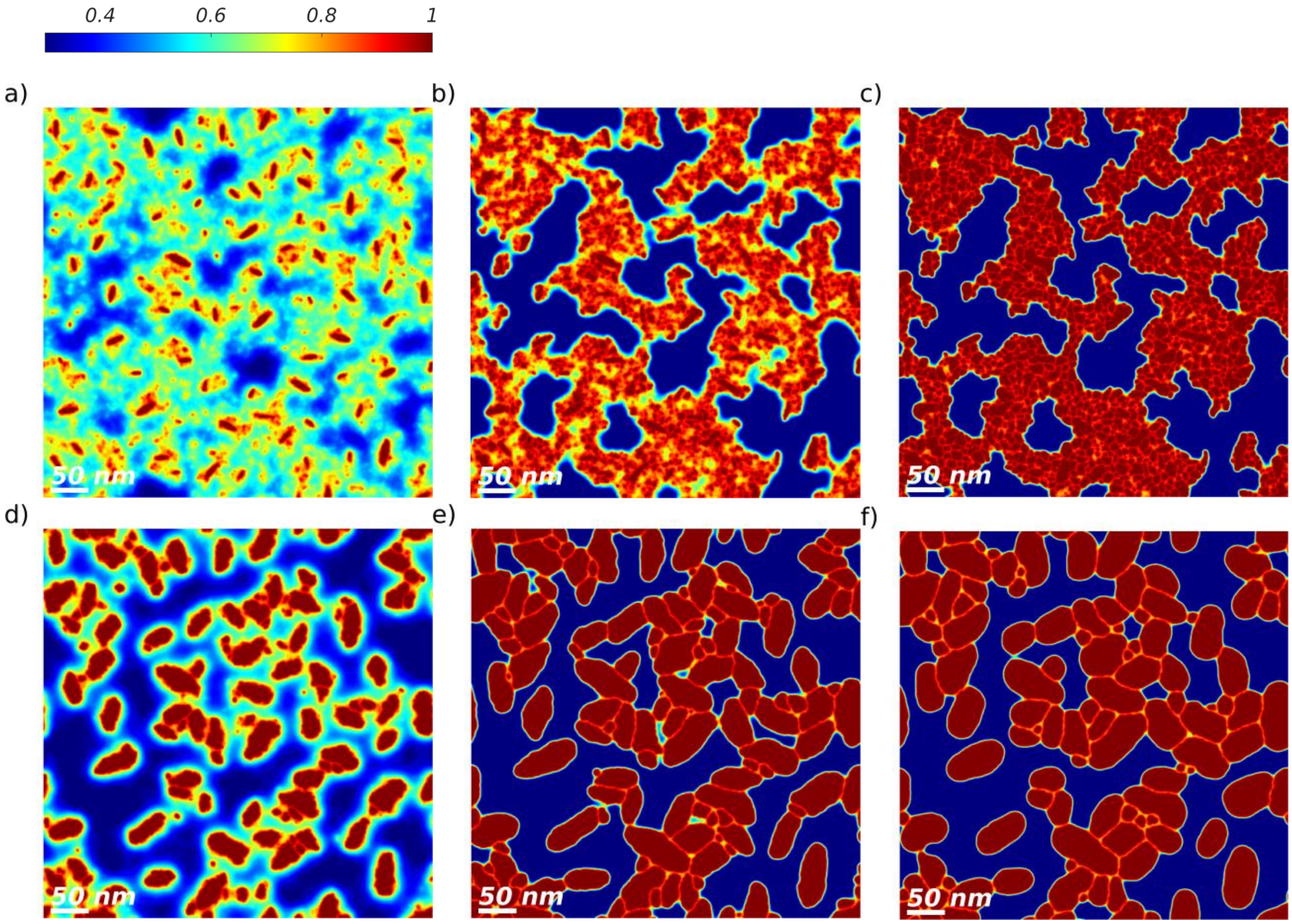

*Figure 7: TA at 80 $°C$ (top row) and 140 $°C$ (bottom row) of a binary DRCN5T:$PC_{71}$BM blend with initial blend ratio 1:0.8, simulation of type C. The DRCN5T volume fraction fields are shown, after TA of 150 $s$ (a and d), 300 $s$ (b and e), and 600 $s$ (c and f). The volume fraction colorbar spreads from dark blue (Low DRCN5T content) to dark red (pure DRCN5T).*

## 3.4. Amorphous-Amorphous phase separation

AAPS is another possible physical phenomenon for the morphology evolution of the DRCN5T:$PC_{71}$BM film. In this section, we trigger AAPS by choosing a value of $\chi_{aa}$ above the critical value (simulations of type D). The objective is to determine whether BHJ in which AAPS takes place have morphologies in line with the experimental observations of Sec. 2.1. This will allow to conclude whether AAPS is active during the thermal annealing of the DRCN5T:$PC_{71}$BM active layer. The morphology at a TA temperature of 80 $°C$ is shown in Figure 8. As a result of donor-acceptor immiscibility, the amorphous phase demixes into a donor and an acceptor amorphous phase. As a result, there are 3 phases in the BHJ (Figure 8b). The first phase is the very pure crystalline DRCN5T material, the second is the amorphous $PC_{71}$BM-rich phase, and the third one is the amorphous DRCN5T-rich phase (See Sec. 8 in SI). However, in the EFTEM image, there are only 2 phases at the final stage of the TA: the amorphous mixed phase and the crystalline DRCN5T material. The mismatch between simulations of type D and the experimental observations allows to conclude that AAPS is not triggered during TA of the DRCN5T: $PC_{71}BM$ active layer.

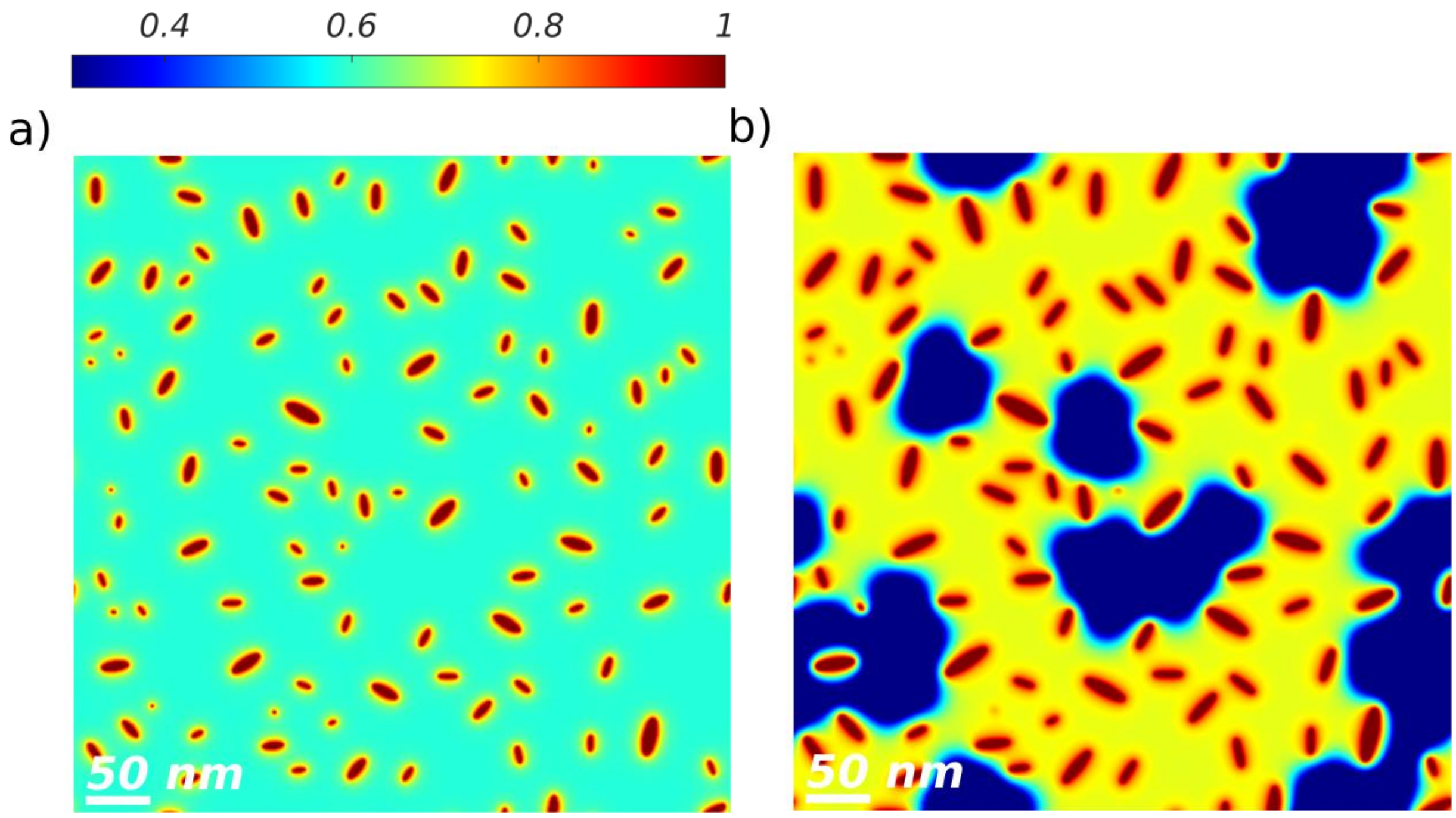

*Figure 8: TA at 80 $°C$ of a binary DRCN5T:$PC_{71}BM$ blend with initial blend ratio 1:0.8, simulation of type D. The DRCN5T volume fraction fields are shown for the as-cast film with low crystals density (a) and at TA time of 600 $s$ (b). The volume fraction colorbar spreads from dark blue (Low DRCN5T content) to dark red (pure DRCN5T).*

## 3.5. Diffusion-limited crystal growth

The remaining question to be answered to is whether crystal growth is diffusion-limited during the TA of the studied all small-molecule system. To this end, simulations with diffusion-limited crystal growth (type E) are performed and the resulting morphologies compared to the experimental observations of Sec. 2.1. The morphologies after 10 $min$ of TA at temperatures of 80 $°C$, 100 $°C$, 120 $°C$, 140 $°C$ and 160 $°C$ are shown in Figure 9. Provided the growth process is sufficiently fast and advanced (120 $°C$ and beyond), depletion zones characteristic of the diffusion-limited regime form around the DRCN5T crystals (Figure 9c-f). The DRCN5T volume fraction in the depletion zones becomes lower at higher temperatures. This is due to the larger and faster crystal growth with increasing temperature. Notably, the leaf shape of the DRCN5T crystals is well conserved. Apart from the thin depletion areas around the crystals, the morphologies are similar to those observed without diffusion limitation (compare with Figure 3). Experimentally, the question of the presence of depletion zones around the DRCN5T crystals during the TA remains difficult to solve, even though Wu and coworkers have identified an enrichment of $PC_{71}BM$ around the DRCN5T crystals using 4D scanning confocal electron diffraction (4D-SCED) for a TA of 8 $min$ at 140 $°C$.[21] On the one hand, this means that the occurrence of diffusion limitations during crystal growth cannot be fully confirmed or discarded on basis of the comparison between simulation results (type A and E) and the experimental observations. On the other hand, this shows that the influence of diffusion limitations on morphology is anyway very limited, which means that diffusion limitations do not represent a decisive driver for morphology development.

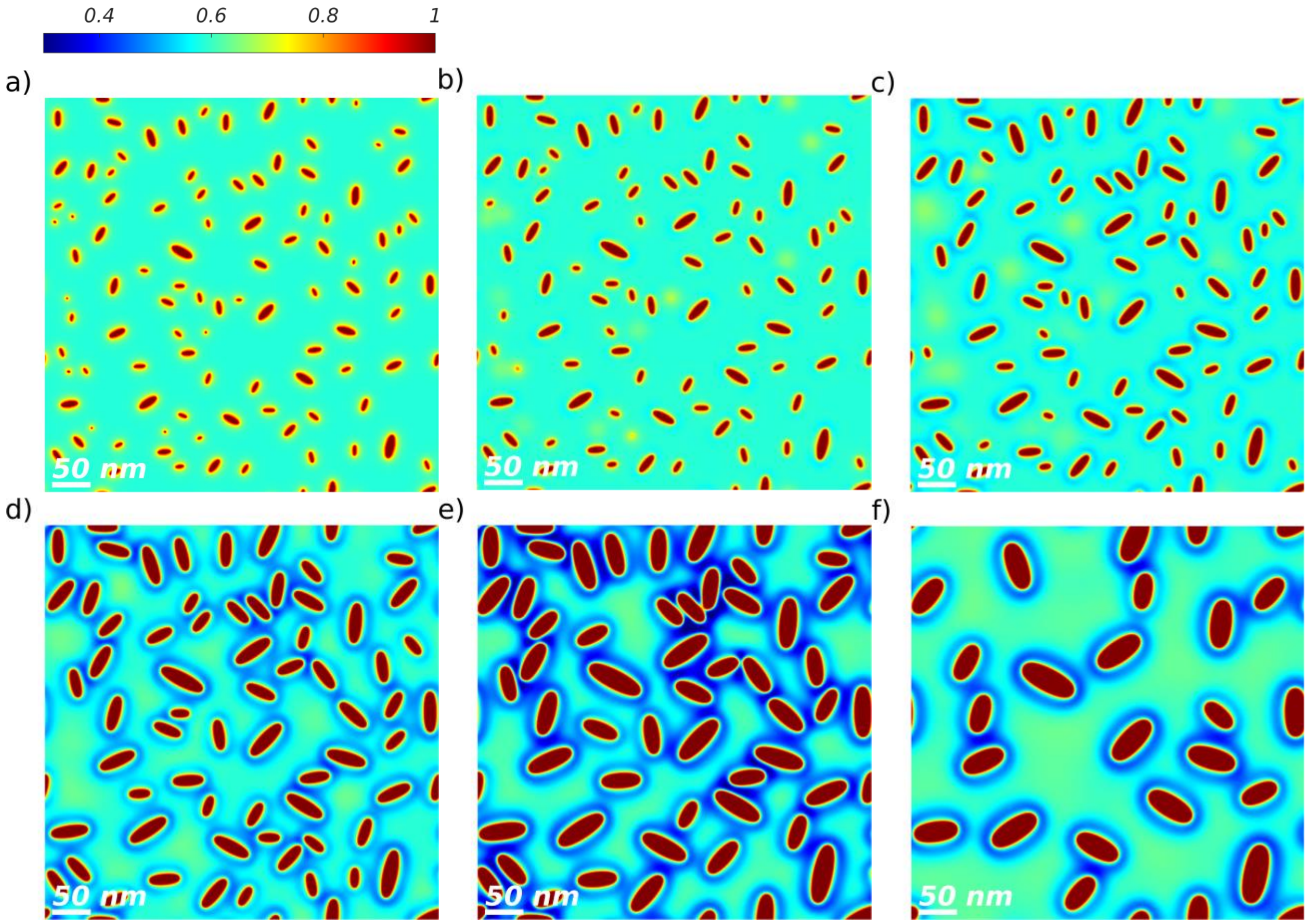

*Figure 9: TA of a binary DRCN5T: $PC_{71}BM$ blend with initial blend ratio 1:0.8, simulation of type E. The DRCN5T volume fraction fields are shown for the as-cast film with low crystal density (a) and after 600 $s$ at TA temperatures of 80 $°C$ (b), 100 $°C$ (c), 120 $°C$ (d), 140 $°C$ (e), and 160 $°C$ (f). The crystal growth is diffusion-limited. The volume fraction colorbar spreads from dark blue (Low DRCN5T content) to dark red (pure DRCN5T).*

# 4. Conclusion and outlook

We applied a phase field model to simulate the TA of a DRCN5T:$PC_{71}BM$ film, mimicking the experiments conducted in Ref. [8,21]. The model parameters were adjusted based on the experimental values published for the DRCN5T:$PC_{71}BM$ system. Physics-based temperature dependencies were used for the material parameters. The objective was to determine which physical processes are the main drivers of the BHJ morphology evolution under TA. From the comparison of the simulation results with experimental data,[8] we conclude that the morphology evolution during TA is mainly driven by (1) the early dissolution of the smallest as-cast crystals that become thermodynamically unstable at higher temperatures and (2) the growth of the remaining large crystals fed by the small molecules available in the amorphous matrix. The simulation results do not provide any conclusive answer regarding a diffusion-limited regime during crystal growth, since it only has a very limited qualitative impact on the morphology. All other possible physical processes investigated in this work (AAPS, nucleation, crystal impingement, grain coarsening, Ostwald ripening) can be considered inactive or negligible. Impingement, as a physical phenomenon leading to crystals deviating from the initial leaf shape, remains untriggered in the present case. AAPS would result in the formation of a three-phase morphology, which is not observed in the measurements. Nucleation would increase the number of crystals upon annealing, which is clearly not observed either. Ostwald ripening and grain coarsening are late-stage processes that are not involved in the early crystal dissolution and hardly influence the late morphology evolution within the investigated TA times. Table 3 summarizes these conclusions on the various considered physical processes.

| Phenomenon | BHJ morphology evolution driver? |
|---|---|
| **Crystal nucleation** | **No** |
| **Grain coarsening** | **No** |
| **Impingement** | **No** |
| **AAPS** | **No** |
| **Diffusion limitation** | **Not conclusive** |
| **Growth of large crystals** | **Yes** |
| **Instability of small crystals** | **Yes** |
| **Ostwald ripening** | **No** |

*Table 3: Summary of the influence of the simulated physical processes on the DRCN5T:$PC_{71}BM$ BHJ evolution upon TA*

SVA treatments were also done on the same DRCN5T:$PC_{71}BM$ samples,[8] which led to PCEs comparable to the one obtained with TA. Furthermore, the SIMS measurements realized by Min and coworkers on SVA and TA show a significant microstructure evolution in the SVA morphology after aging,[18,20] as well as a drop in PCE. In the future, the phase field approach presented here for the investigation of TA can also be used to understand the morphology evolution after SVA. More generally, it can be applied to understand the formation of the BHJ morphology upon processing and its stability upon thermal loading. Moreover, further studies on the morphology-performance relationship will allow to investigate how the efficiency and the intrinsic stability of OSC depend on the BHJ morphology, and therefore on the fabrication procedures.

# 5. Conflict of interest

There are no conflicts to declare.

# 6. Data availability

The simulation data used for this article is publicly accessible (see DOI https://zenodo.org/doi/10.5281/zenodo.14267943).

# 7. Acknowledgements

The authors acknowledge financial support by the German Research Foundation (DFG, Project HA 4382/14-1 and SFB 953), the European Commission (H2020 Program, Project 101008701/ EMERGE), and the Helmholtz Association (SolarTAP Innovation Platform).

# Supporting information

# Phase field simulations of thermal annealing for all-small molecule organic solar cells

Yasin Ameslon[1], Olivier J. J. Ronsin[1], Christina Harreiß[2], Johannes Will[2], Stefanie Rechberger[2], Mingjian Wu[2], Erdmann Spiecker[2] and Jens Harting[1,3]

[1]Helmholtz Institute Erlangen-Nürnberg for Renewable Energy, Forschungszentrum Jülich, Fürther Strasse 248, 90429 Nürnberg, Germany

[2]Institute of Micro- and Nanostructure Research (IMN) & Center for Nanoanalysis and Electron Microscopy (CENEM), Interdisciplinary Center for Nanostructured Films (IZNF), Department of Materials Science and Engineering, Friedrich-Alexander-Universität Erlangen-Nürnberg, Cauerstrasse 3, 91058 Erlangen, Germany

[3]Department of Chemical and Biological Engineering and Department of Physics, Friedrich-Alexander-Universität Erlangen-Nürnberg, Cauerstrasse 1, 91058 Erlangen, Germany

## Contents

# 1. Anisotropy functions

Several anisotropy $\pi$-periodic functions have been tested in order to evaluate the impact of such a choice on the shape of the DRCN5T crystals. They are defined on $[0,\pi]$, as follow:

$$f_1(\theta) = \frac{cos(2\theta) + 1}{2} \quad \text{(S1)}$$

$$f_2(\theta) = 1 \text{ when } \theta < \frac{\pi}{4} \text{ and } \pi > \theta > \frac{3\pi}{4}, \text{ and } f_2(\theta) = 0 \text{ otherwise} \quad \text{(S2)}$$

$$f_3(\theta) = |cos(\theta)| \quad \text{(S3)}$$

$$f_4(\theta) = \frac{e^{\theta-\frac{\pi}{2}-1}}{e^{\frac{\pi}{2}-1}} \text{ when } \pi > \theta > \frac{\pi}{2}, f_4(\theta) = \frac{e^{-\theta+\frac{\pi}{2}-1}}{e^{\frac{\pi}{2}-1}} \text{ when } \theta < \frac{\pi}{2} \quad \text{(S4)}$$

$$f_5(\theta) = \frac{\left(\frac{\pi}{2} - \theta\right)^2}{\frac{\pi^2}{2}} \quad \text{(S5)}$$

The anisotropy functions are illustrated in Figure S1.

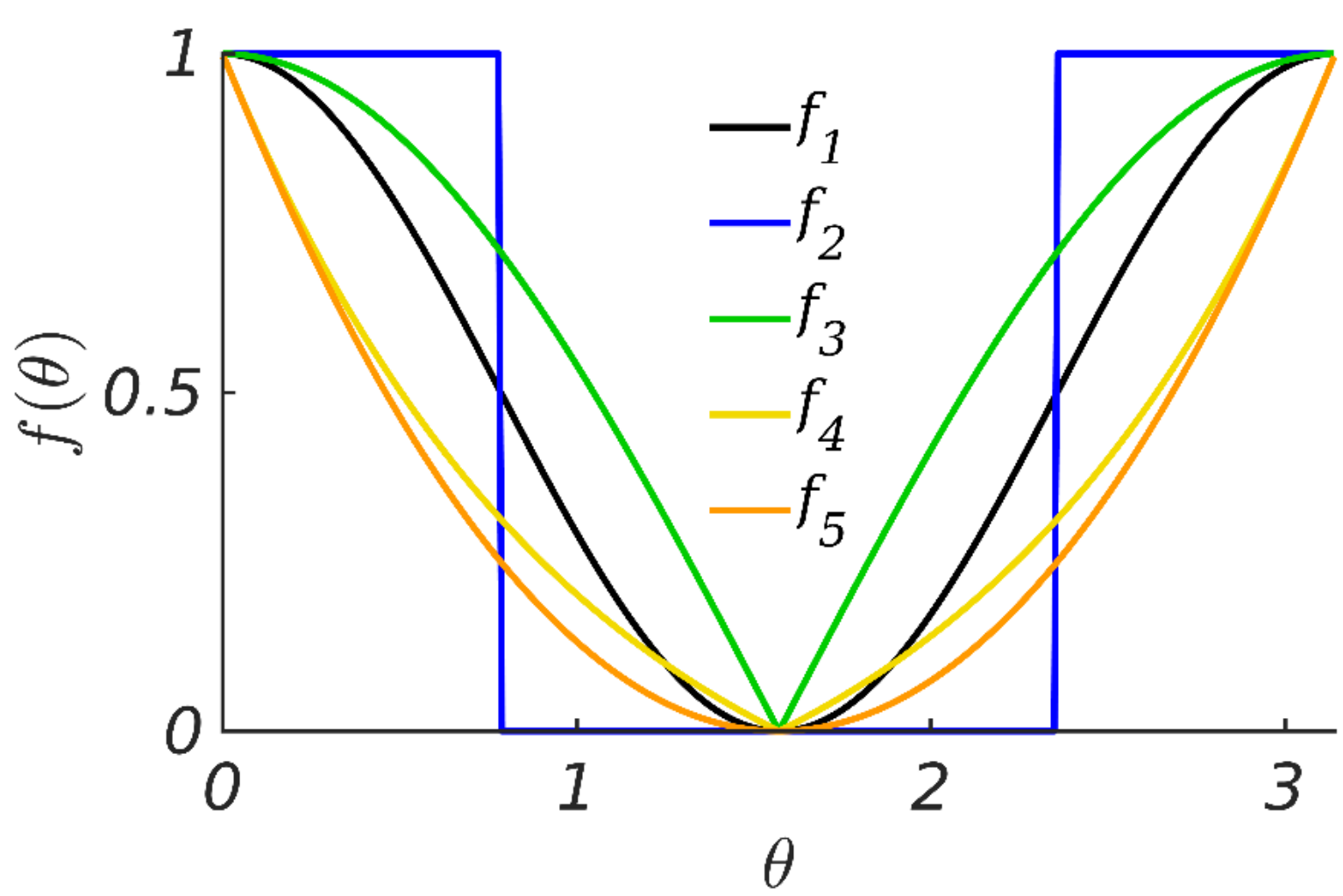

*Figure S1: Anisotropy functions tested to obtain the DRCN5T crystals leaf shape.*

The anisotropy functions are implemented on the AC mobility as described in Sec. 2 of the main text. They are tested by letting an initially elliptic crystal grow. The shapes obtained after significant growth are illustrated using the order parameter field $\phi$ in Figure S2 (a), (b), (c), (d), and (e) for the anisotropy functions $f_1$, $f_2$, $f_3$, $f_4$, and $f_5$, respectively.

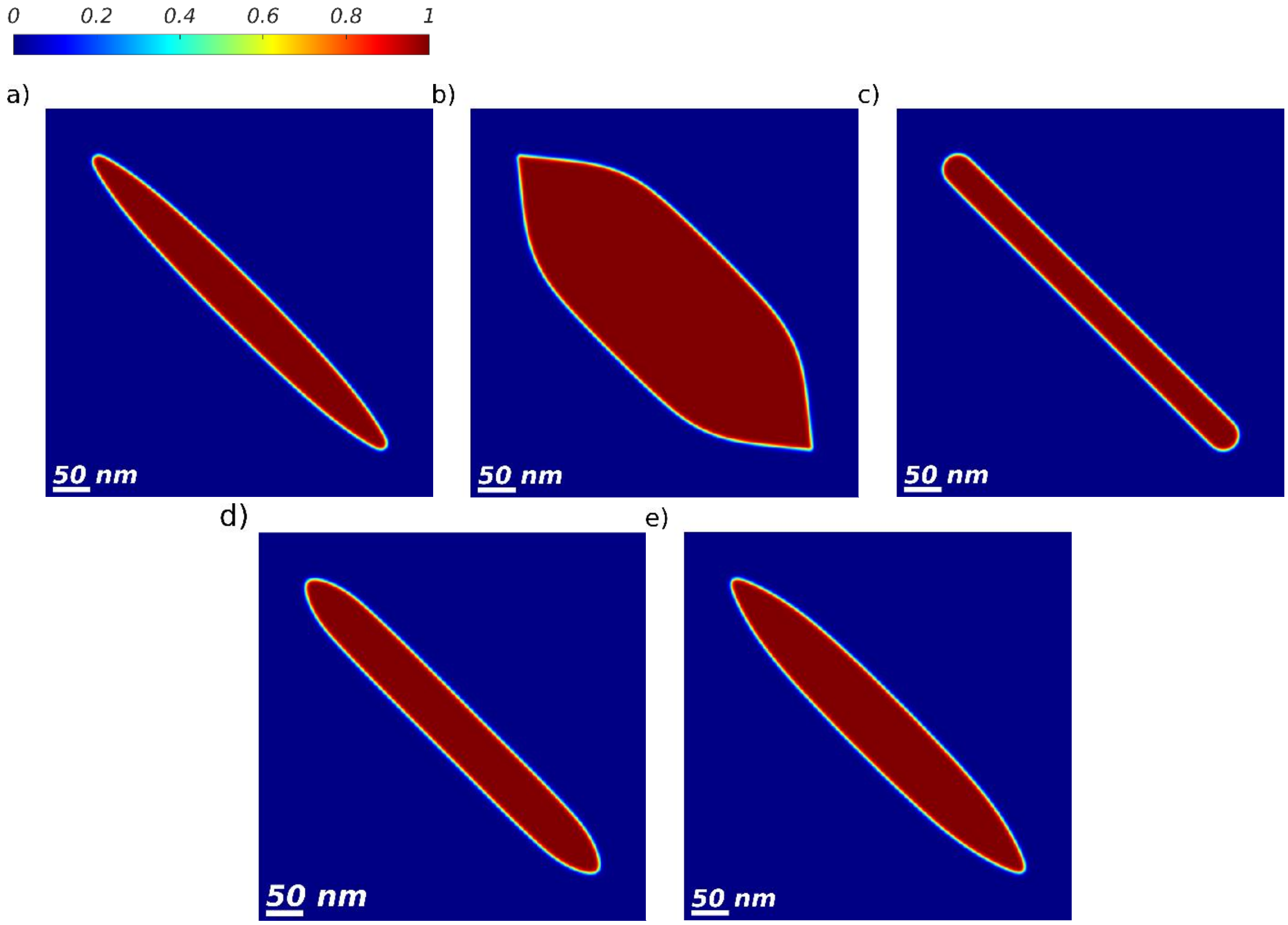

*Figure S2: Order parameter field after pure crystal growth with different anisotropy functions. The color scales from dark blue (amorphous material) to dark red (crystalline material).*

First of all, only functions $f_1$ and $f_5$ generate leaf shaped crystals similar to the ones observed in the EFTEM images obtained by Harreiß and coworkers.[1] Moreover, the longitudinal-to-transversal growth rate ratios extracted from the simulation results are compared to the expected values $M_{ratio} + 1$ values for the different functions. The results are summarized in Figure S3. The expected longitudinal-to-transversal growth rate ratio is better recovered with function $f_1$ than with function $f_5$ and is accurately recovered up to ratios as large as 4 with function $f_1$. Because of its physical meaning, of the proper reproduction of the leaf shape and of the growth ratios $M_{ratio} + 1$, the function $f_1$ is chosen to simulate the anisotropic growth of the DRCN5T crystals.

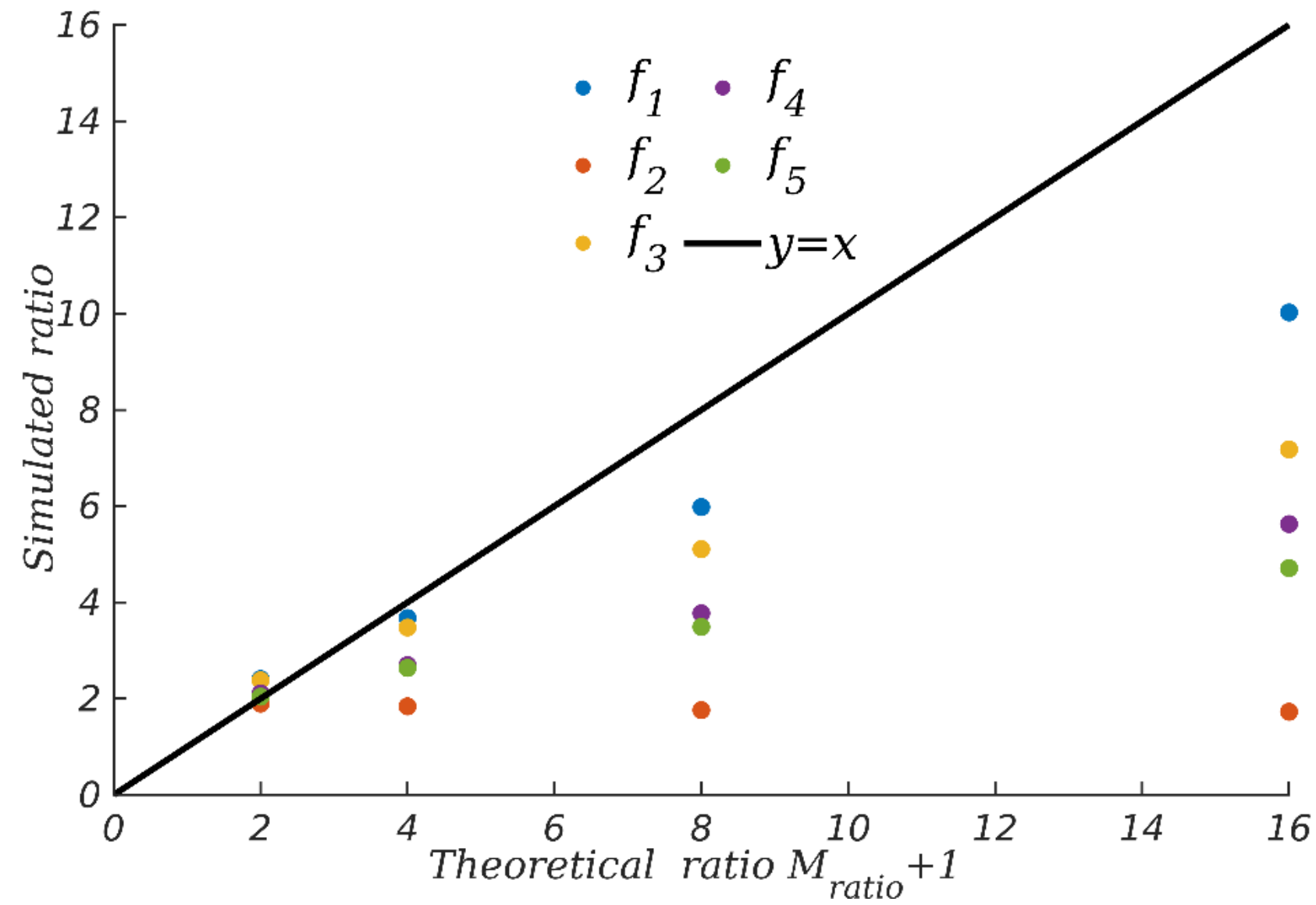

*Figure S3: Longitudinal-to-transversal growth rate ratios extracted from the simulations, compared to the expected theoretical ones, for different anisotropy functions.*

## 2. Determination of the amorphous-amorphous interaction parameter

The melting point depression of DRCN5T was measured by Zhang and coworkers for different blend ratios of DRCN5T:$PC_{71}BM$ [2]. The data is reused to calculate the value $\chi_{aa}(T_m)$ using the melting point depression equation below:

$$\frac{1}{{T_m}^{blend}(\varphi)} - \frac{1}{T_m} = -\frac{Rv_2}{L_{fus}v_1}\left[\frac{\ln(\varphi)}{N_1} + \left(\frac{1}{N_1} - \frac{1}{N_2}\right)(1-\varphi) + \chi_{aa}(T_m)(1-\varphi)^2\right] \quad \text{(S6)}$$

The amorphous-amorphous interaction parameter is assumed to be constant for temperatures close to the melting point temperature $T_m$ of a pure DRCN5T crystal. ${T_m}^{blend}$ is the melting temperature of DRCN5T crystalline material in a DRCN5T: $PC_{71}BM$ blend[3] and $v_1$ is the molar volume of DRCN5T. The linear regression of $\frac{1}{{T_m}^{blend}(\varphi)} - \frac{1}{T_m} + \frac{Rv_2}{L_{fus}v_1}(\frac{\ln(\varphi)}{N_1} - \left(\frac{1}{N_1} - \frac{1}{N_2}\right)(1-\varphi)$ as a function of $(1-\varphi)^2$ gives the value $\chi_{aa}(T_m) = 1.4$.

The critical interaction parameter, below which the mixture is miscible, is $\chi_{crit} = 1.54$. For simulations of type D, whereby an AAPS is desired (See Sec. 2 in the main text), the interaction parameter is chosen above the critical value at the simulated TA temperatures. For all other simulation types the interaction parameter is set to be below the critical value at the simulated TA temperatures, so that there is no AAPS. The two phase diagrams corresponding to both cases, i.e. $\chi \leq \chi_{crit}$ (Figure S4a) and $\chi > \chi_{crit}$ (Figure S4b) are shown in Figure S4.

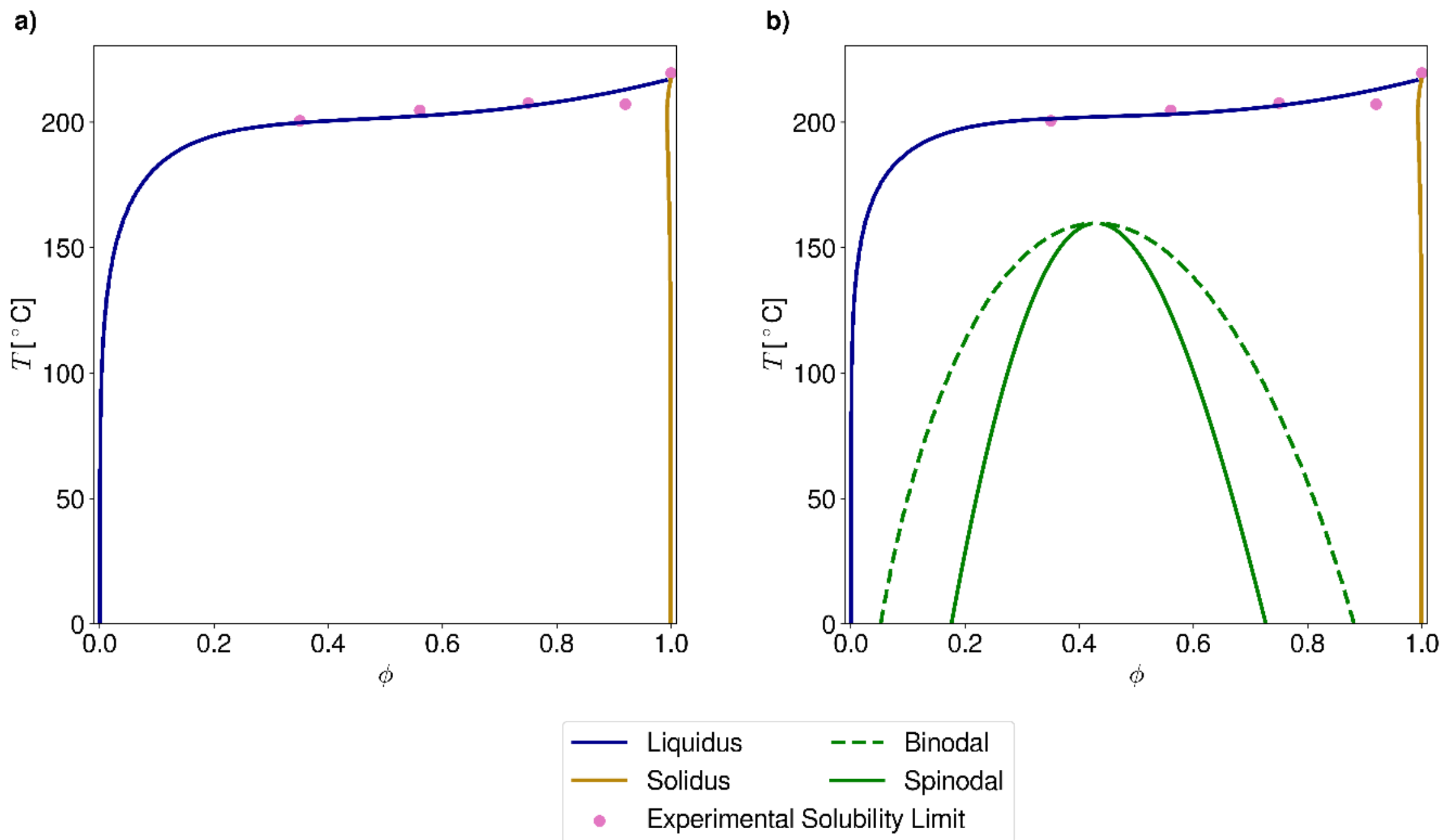

*Figure S4: Phase diagrams of the studied DRCN5T:$PC_{71}BM$ mixture assuming miscibility (a) and immisciblility (b). The liquidus line is in blue, the solidus in yellow, the fitted experimental solubility limit points in pink. The increase of $\chi_{aa}$ from $\chi_{aa}(80°C) = 1.1$ (for a miscible blend) to $\chi_{aa}(80°C) = 1.8$ (for an immiscible blend) results in the appearance of the unstable and metastable regions, up to high temperatures (150°C) centered around $\varphi = 0.4$*

# 3. Temperature-dependence of the critical radius and nucleation rate derived from classical nucleation theory

In the sense of the classical nucleation theory (CNT), a germ must reach a critical size in order to be stable and grow, due to the competition between the driving force for crystallisation and the surface energy $\sigma$. The expression of the free energy in 2D can be written as follow for a crystal germ of radius $r$ :

$$G = \pi r^2 \rho L_{fus}(\frac{T}{T_m} - 1) + 2\pi r\sigma \tag{S7}$$

The critical radius $r^*$ corresponds to the value of the radius for which the free energy derivative is equal to 0:

$$r^* = \frac{\sigma}{\rho L_{fus}(1 - \frac{T}{T_m})} \tag{S8}$$

Using the expression of the surface energy $\sigma = \frac{\varepsilon\sqrt{\rho W_{fus}}}{3\sqrt{2}}$ derived from the Allen-Cahn equation,[4] the critical radius can be expressed depending on the PF model parameters as:

$$r^* = \frac{\varepsilon\sqrt{W_{fus}}}{3L_{fus}(1-\frac{T}{T_m})\sqrt{2\rho}} \quad \text{(S9)}$$

In the CNT, the nucleation rate is $v_{nucl} \propto e^{\frac{-G^*}{RT}}$ where $G^*$ is the energy barrier to be overcome for stable germs to grow.[6] Using S7 and S8 and the expression of the surface energy in the sense of the phase field parameters, the following expression of $G^*$ is obtained:

$$G^* \sim \frac{\varepsilon^2 W_{fus}}{L_{fus}(1-\frac{T}{T_m})} \quad \text{(S10)}$$

The temperature dependency $\varepsilon \propto \sqrt{T}$ and $W_{fus} \propto T$ described in Sec. 2 of the article can be used with equation S10 and the expression of the nucleation rate $v_{nucl} \propto e^{\frac{-G^*}{RT}}$ to obtain the following temperature dependency for the nucleation rate in a pure material:

$$v_{nucl} \propto e^{\frac{-\alpha_{nucl}T}{(1-\frac{T}{Tm})}} \quad \text{(S11)}$$

where $\alpha_{nucl}$ is a positive constant. The nucleation rate consequently decreases with increasing temperature.

# 4. Determination of the mobility coefficient by fitting the experimental growth rates

The longitudinal and transversal growth rates were calculated from the crystals lengths and widths measured by Harreiß and coworkers[1] in the as-cast film and in the thermally annealed films. The calculated growth rates at the different TA temperatures are summarized in Table S1.

| TA temperature | Longitudinal growth rate (nm/s) | Transversal growth rate (nm/s) |
|---|---|---|
| **80 °C** | **0.0193** | **0.0032** |
| **100 °C** | **0.0217** | **0.0052** |
| **120 °C** | **0.0245** | **0.0070** |
| **140 °C** | **0.0610** | **0.0155** |
| **160 °C** | **0.1662** | **0.0402** |

*Table S1: experimentally measured longitudinal and transversal growth rates of the DRCN5T crystals at different TA temperatures*[1]

In the Allen-Cahn model, the growth rate of a pure crystal $v$ is such that $v \propto \frac{1}{\sqrt{W}}\varepsilon M_{aniso}(1-\frac{T}{T_m})$.[4] Assuming that the same temperature dependency will still hold in a blend, and inserting the physical temperature dependencies of $\varepsilon$, $M_{aniso}$ and $W$ described in Sec. 2 of the main text, we obtain $v \propto e^{\frac{-\beta_D}{T}}(1-\frac{T}{T_m})$,. A linear fit of $ln(v) - ln(1-\frac{T}{T_m})$ against $\frac{1}{T}$ is done to determine the value of $\beta_D$. Note that Harreiß and coworkers[1] assumed $v \propto e^{\frac{-\beta_D}{T}}$. A linear fit of $ln(v)$ against $\frac{1}{T}$ is done additionally to compare the differences between the 2 methods. The linear fits for the longitudinal and transversal growth rates are shown in Figure S5. There is no significant difference between the 2 methods previously described as observed in Figure S5.

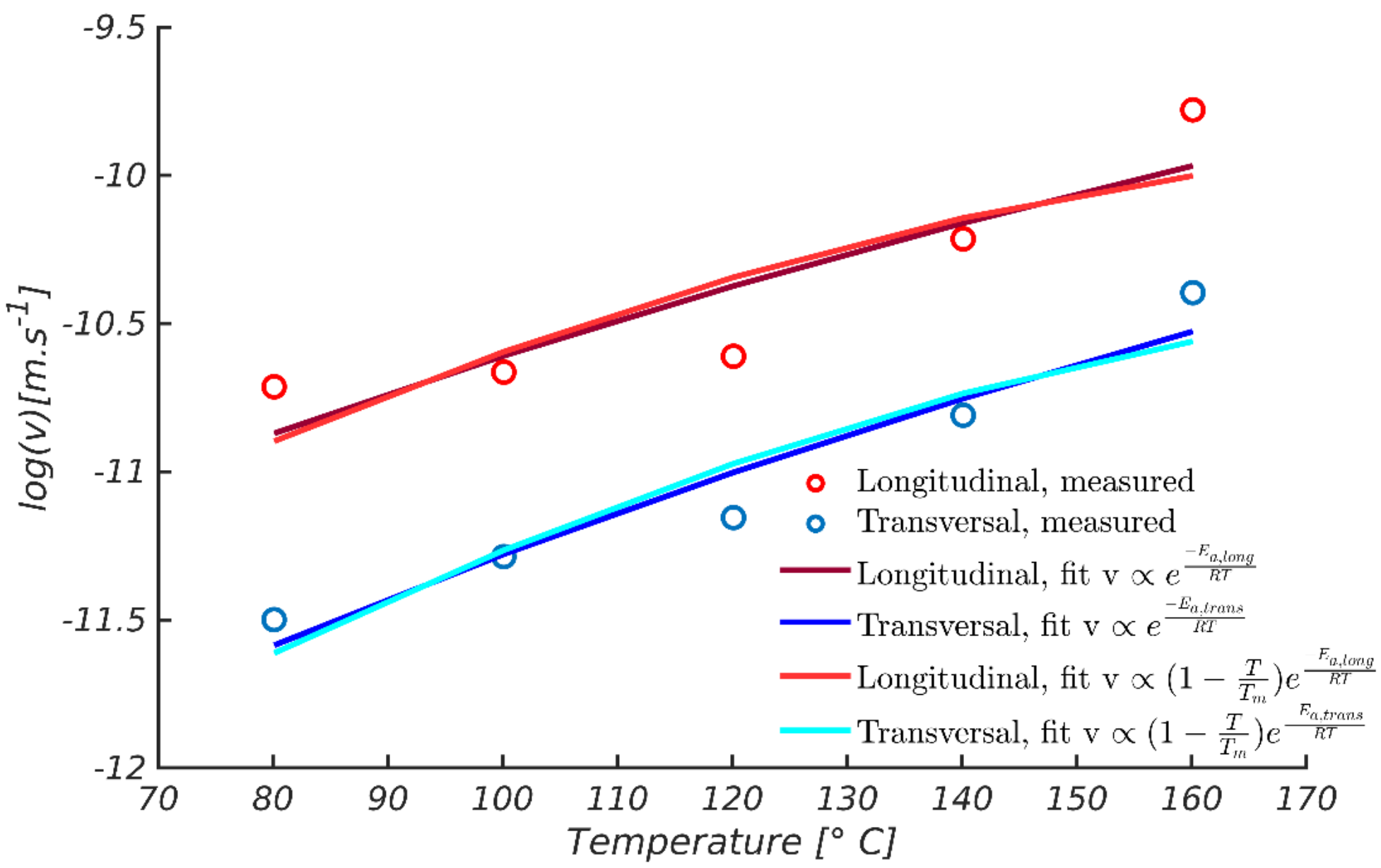

*Figure S5: Arrhenius fit of the transversal and longitudinal growth rates.*

Moreover, the corresponding longitudinal growth activation energy is found to be $E_{a,long} = 47$ kJ, and the transversal growth activation energy $E_{a,trans} = 52$ kJ. On the one hand, this longitudinal growth activation energy was used for parametrizing the temperature dependence of the self-diffusion coefficients $D_{ij}$ and the mobility $M_0$ coefficients described in Sec. 2 of the main text. On the other hand, the longitudinal to transversal growth rates ratio $M_{ratio} + 1$ described in Sec. 2 of the main text is chosen such that $M_{ratio} + 1 \propto e^{\frac{E_{a,trans} - E_{a,long}}{RT}}$.

# 5. Full list of simulation parameters values

| Parameters | Full name | Values/Temperature dependency ($T$ in $K$) | Unit |
|---|---|---|---|
| $\rho$ | DRCN5T density | $1267$ | $kg\ m^{-3}$ |
| $W_{fus}$ | Energy barrier upon crystallisation | $108.48 * 10^{-3} * T$ | $kJ\ kg^{-1}$ |
| $L_{fus}$ | DRCN5T heat of fusion | $44.89$ | $kJ\ kg^{-1}$ |
| $T_m$ | DRCN5T melting temperature | $490.25$ | $K$ |
| $N_1, N_2$ | Species molar size | $1.75, 1$ | - |
| $v_0$ | **Smallest component molar volume** | $5.69 * 10^{-4}$ | $m^3\ mol^{-1}$ |
| $\chi_{aa}$ | **Amorphous-amorphous interaction parameter** | **Simulations of type A, B, C and E**<br>$2.17 - 379.23 * \frac{1}{T}$<br>**Simulations of type D**<br>$0.37 + 505.64 * \frac{1}{T}$ | - |
| $\chi_{ca}$ | **Crystalline-amorphous interaction parameter** | $1349.8 * \frac{1}{T}$ | - |
| $\varepsilon_g$ | **Surface tension parameter for grain boundaries** | **0.06** | $J\ m^{-2}$ |
| $\varepsilon$ | **Surface tension parameter for crystallinity order parameter gradient** | $5.85 * 10^{-7} * \sqrt{T}$ | $(J\ m^{-1})^{0.5}$ |
| $\kappa$ | **Surface tension parameter for composition gradients** | $2 * 10^{-10}$ | $J\ m^{-1}$ |
| $D_{11}, D_{12}, D_{21}, D_{22}$ | **Self diffusion coefficients** | **Simulations of type A, B, C, and D**<br>$D_{11} = D_{21} = 3.19 * 10^{-9} * T * e^{-\frac{5621.9}{T}}$<br>$D_{12} = D_{22} = 6.37 * 10^{-9} * T * e^{-\frac{5621.9}{T}}$<br>**Simulations of type E**<br>$D_{11} = D_{21} = 7.35 * 10^{-16} * T * e^{-\frac{5621.9}{T}}$<br>$D_{12} = D_{22} = 1.47 * 10^{-15} * T * e^{-\frac{5621.9}{T}}$ | $m^2\ s^{-1}$ |
| $M_0$ | **Crystal mobility coefficient** | $18.31 * T * e^{-\frac{5621.9}{T}}$ | $s^{-1}$ |
| $M_{ratio} + 1$ | **Longitudinal/transversal growth ratio** | $1.59 * e^{\frac{688.3}{T}}$ | - |

*Table S2: Simulation parameters used for the simulations presented in the paper.*

# 6. Temperature-dependent crystals stability in the DRCN5T-PCBM blend

To obtain the temperature-dependent critical size of DRCN5T germs in the DRCN5T-PCBM mixture, simulations with initially only one single elliptic crystal are performed. Thereby, the simulations parameter set is identical to the one of simulations of type A (See Sec. 2 of the main text), so that the only possible evolution is dissolution (if unstable) or growth (if stable) of the initial crystal. In order to determine the critical crystal size above which the crystals grow, the initial crystal size is varied (however with a fixed major-to-minor axis length ratio based on the DRCN5T crystals lengths and widths measured in the as cast film by Harreiß and coworkers[1,5]). Observing whether the crystal disappears or grows, provides the temperature-dependent critical size.

Once this critical size is known, since the size distribution of the germs in the as-cast film is also known, the number of crystals expected to be unstable upon TA is also known. This number is compared to the number of crystals really dissolved at the end of the 10 min of TA in simulations of type A. As illustrated in Figure S6, the number of crystals really dissolved is coherent with the expectations. Together with the fact that these crystals disappear at the beginning of annealing, we conclude that the crystals dissolution upon TA is due to their instability and not to Ostwald ripening.

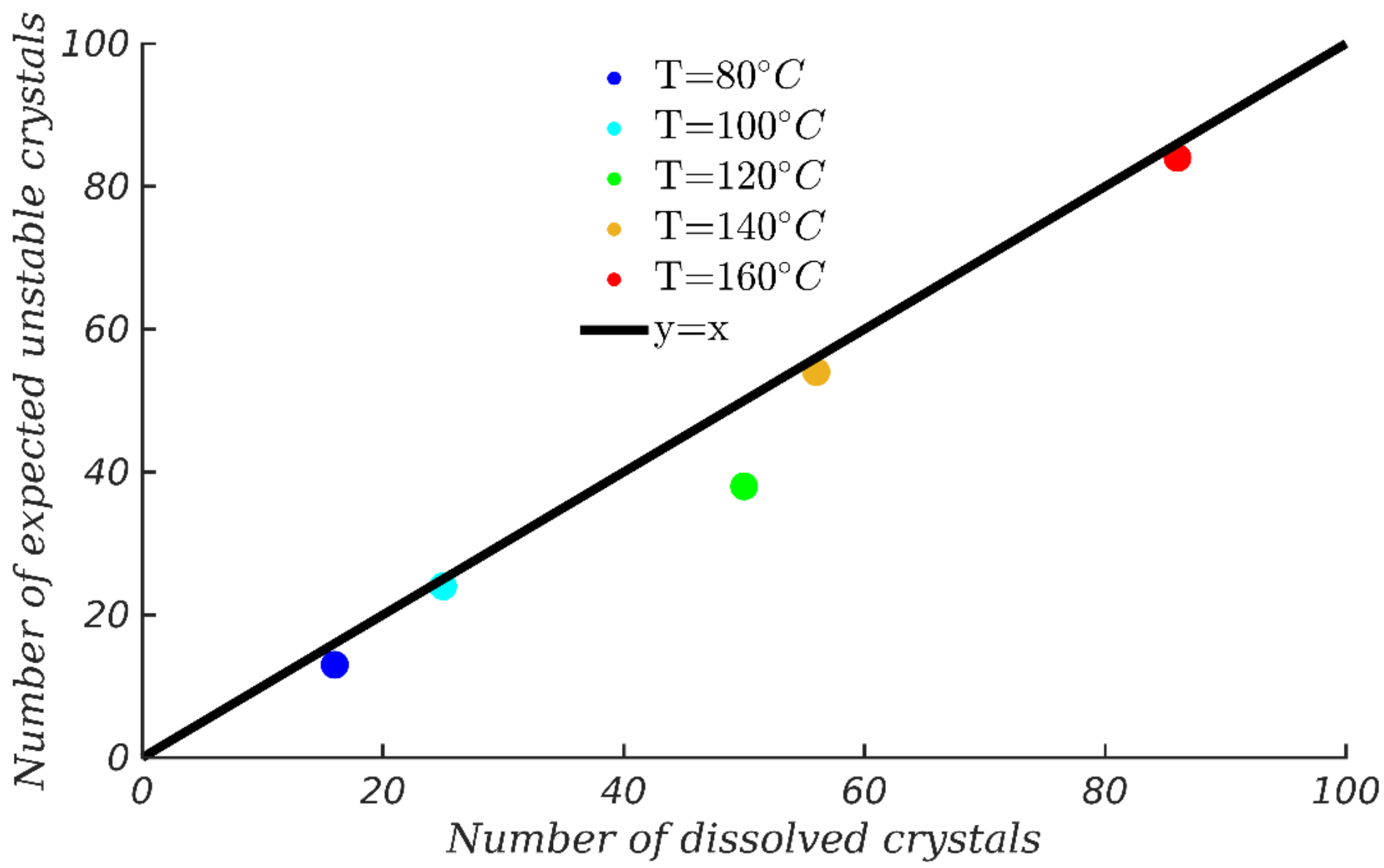

*Figure S6: Number of expected unstable initial crystals as a function of the number of dissolved crystals for simulations of Type A after 10 min of TA at temperatures of 80°C, 100°C, 120°C, 140°C, and 160°C. The black curve corresponds to the y=x line.*

# 7. Impact of coarsening during TA for simulations with nucleation and coarsening (type C)

Grain coarsening becomes generally dominant and visible only at late-stage after nucleation and crystal growth. The DRCN5T crystallinity $\chi_{crys}$ curve and number of crystals is determined for simulations of type C at 140°C (see main text), as illustrated in Figure S7. The crystallinity increases from $\chi_{crys} \approx 0.05$ to $\chi_{crys} \approx 0.9$ and the number of crystals from 40 to 130 between $t = 0s$ and $t = 250\ s$ during the nucleation and growth stage. After $t \approx 250\ s$, the effect of nucleation and growth is negligible compared to coarsening, which is reflected after $t \approx 250\ s$ on the crystallinity curve by a decrease of the curve slope and reflected as well by a decrease in the number of crystals.

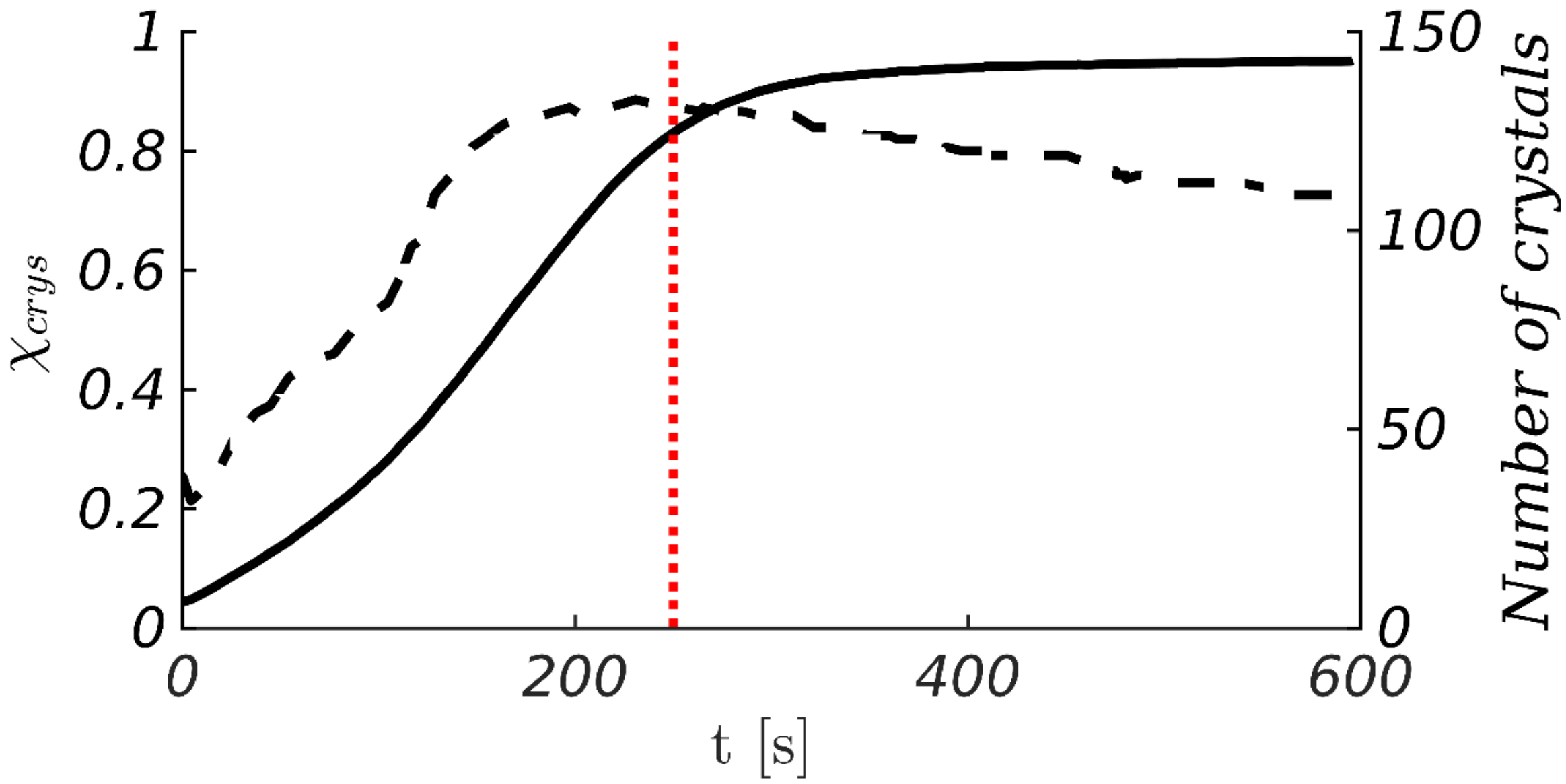

*Figure S7: Evolution of the proportion of crystalline DRCN5T (black solid line and left y-axis) and number of DRCN5T crystals (black dashed line and right y-axis). The red dotted line delimits the nucleation and growth, and coarsening stages.*

## 8. Order parameter and orientation plot for simulations with amorphous-amorphous phase separation (type D)

The order parameter fields $\phi$ and orientation fields $\theta$ allow to track the crystalline regions and the crystals orientation. As illustrated in Figure S8a,b the high-volume fraction phase ($\varphi \approx 1$) is crystalline. In simulations of type D, both other phases with DRCN5T volume fractions of $\varphi \approx 0.7$ and $\varphi \approx 0.1$ are amorphous. The edge-on crystals orientations are shown in Figure S8c.

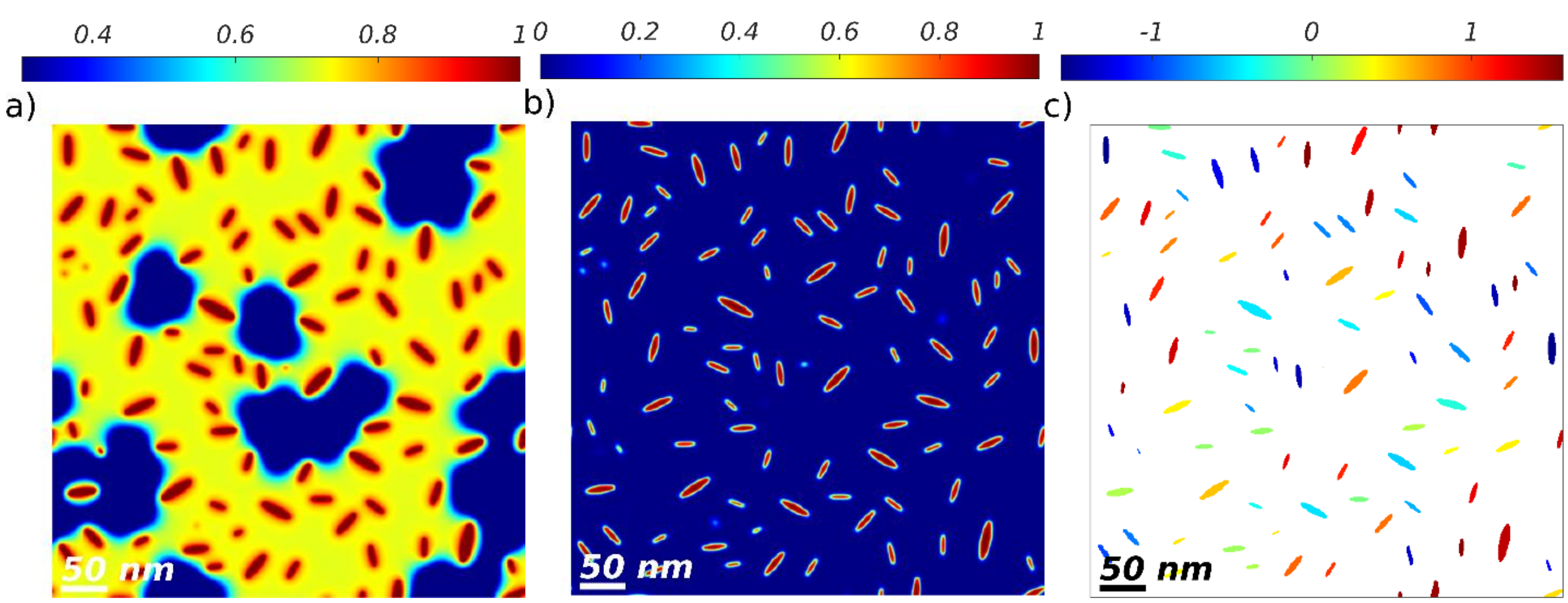

*Figure S8: TA at 80°C of a binary DRCN5T: $PC_{71}BM$ blend with initial blend ratio 1:0.8, simulation of type D. The DRCN5T volume fraction (a), crystalline order parameter (b), and orientation (c) fields are shown for a TA time of 600 s.*